\documentclass[10pt,letterpaper]{article}
\usepackage[top=0.85in,left=2.75in,footskip=0.75in]{geometry}

\usepackage{amsmath,amssymb}

\usepackage{changepage}

\usepackage[utf8x]{inputenc}

\usepackage{textcomp}
\usepackage{marvosym}

\usepackage{cite}

\usepackage{nameref,hyperref}

\usepackage[right]{lineno}

\usepackage{microtype}
\DisableLigatures[f]{encoding = *, family = * }

\usepackage[table]{xcolor}

\usepackage{array}

\newcolumntype{+}{!{\vrule width 2pt}}

\newlength\savedwidth

\usepackage{setspace} 
\raggedright
\usepackage[aboveskip=1pt,labelfont=bf,labelsep=period,justification=raggedright,singlelinecheck=off]{caption}

\makeatletter
\renewcommand{\@biblabel}[1]{\quad#1.}
\makeatother

\usepackage{lastpage,fancyhdr,graphicx}
\usepackage{epstopdf}
\fancyheadoffset[L]{2.25in}
\fancyfootoffset[L]{2.25in}
\usepackage{xspace}
\usepackage{textgreek}
\usepackage{mathtools}
\usepackage{mathrsfs}
\usepackage{enumitem}
\newcommand{\cae}{\textit{Caenorhabditis elegans}\xspace}
\newcommand{\cel}{\textit{C elegans}\xspace}

\newcommand{\petsc}{PETSc\xspace}
\newcommand{\vx}{\ensuremath{\mathbf{x}}}

\newcommand{\vk}{\ensuremath{\mathbf{k}}}
\newcommand{\vv}{\ensuremath{\mathbf{v}}}
\newcommand{\rmax}{\ensuremath{\rho_\text{max}}}

\newcommand{\srhtwo}{\textit{srh--2}\xspace}
\usepackage{indentfirst}
\usepackage{textcomp}
\usepackage{siunitx}
\usepackage{bigints}
\usepackage{physics}
\usepackage{array}
\usepackage[cal=boondox,scr=boondoxo]{mathalfa}
\usepackage{csquotes,xpatch}
\usepackage[english]{babel}
\usepackage{csquotes,xpatch}
\usepackage{fancyhdr}
\newcommand{\beginsupplement}{%
        \setcounter{table}{0}
        \renewcommand{\thetable}{S\arabic{table}}%
        \setcounter{figure}{0}
        \renewcommand{\thefigure}{S\arabic{figure}}%
     }
\begin{document}
\vspace*{0.2in}

% Title must be 250 characters or less.
\begin{flushleft}
{\Large
\textbf\newline{A Keller-Segel model for \cel L1 aggregation} % Please use "sentence case" for title and headings (capitalize only the first word in a title (or heading), the first word in a subtitle (or subheading), and any proper nouns).
}
\newline
% Insert author names, affiliations and corresponding author email (do not include titles, positions, or degrees).
\\
Leon Avery\textsuperscript{1
%\textcurrency
},
Brian Ingalls\textsuperscript{1,*},
Catherine Dumur\textsuperscript{2},
Alexander Artyukhin\textsuperscript{3},
\\
\bigskip
\textsuperscript{\textbf{1}} Department of Applied Mathematics, University of Waterloo, Waterloo, Ontario, Canada
\\
\textsuperscript{\textbf{2}} Department of Pathology, Virginia Commonwealth University, Richmond, VA, USA
\\
\textsuperscript{\textbf{3}} Chemistry Department, State University of New York, College of Environmental Science and Forestry, Syracuse, NY, USA
\\
\bigskip

% Insert additional author notes using the symbols described below. Insert symbol callouts after author names as necessary.
% 
% Remove or comment out the author notes below if they aren't used.
%
% Primary Equal Contribution Note
%\Yinyang These authors contributed equally to this work.

% Additional Equal Contribution Note
% Also use this double-dagger symbol for special authorship notes, such as senior authorship.
%\ddag These authors also contributed equally to this work.

% Current address notes
%\textcurrency Current Address: Dept/Program/Center, Institution Name, City, State, Country
% change symbol to "\textcurrency a" if more than one current address note
% \textcurrency b Insert second current address 
% \textcurrency c Insert third current address

% Deceased author note
%\dag Deceased

% Group/Consortium Author Note
%\textpilcrow Membership list can be found in the Acknowledgments section.

% Use the asterisk to denote corresponding authorship and provide email address in note below.
* bingalls@uwaterloo.ca

\end{flushleft}
% Please keep the abstract below 300 words

%\linenumbers
\newpage

\section*{Abstract}

We describe a mathematical model for the aggregation of starved first-stage \cel larvae (L1s). We propose that starved L1s produce and respond chemotactically to two labile diffusible chemical signals, a short-range attractant and a longer range repellent. This model takes the mathematical form of three coupled partial differential equations, one that describes the movement of the worms and one for each of the chemical signals. Numerical solution of these equations produced a pattern of aggregates that resembled that of worm aggregates observed in experiments. We also describe the identification of a sensory receptor gene, \srhtwo, whose expression is induced under conditions that promote L1 aggregation. Worms whose \srhtwo gene has been knocked out form irregularly shaped aggregates. Our model suggests this phenotype may be explained by the mutant worms slowing their movement more quickly than the wild type.

% Please keep the Author Summary between 150 and 200 words
% Use first person. PLOS ONE authors please skip this step. 
% Author Summary not valid for PLOS ONE submissions.   
\section*{Author summary}

Among the most complex of animal behaviors are collective behaviors, in which animals interact with each other so as to produce large-scale organization. Starved first-stage larvae of the nematode \cae exhibit such a behavior: they come together to form aggregates of several hundred worms. How and why they do this are unknown. To address these questions, we developed a mathematical model of starved L1 aggregation. This model reproduced the main features of the behavior. 

% Use "Eq" instead of "Equation" for equation citations.
\section*{Introduction}

Among the most complex behaviors exhibited by the nematode \cel are social behaviors such as mating \cite{Barr2006} and aggregation. We recently described a new aggregation behavior in starved \cel first-stage larvae (L1s) \cite{Artyukhin2015}. This new behavior raises two broad questions whose answers we lack: (1) How do starved L1s aggregate? I.e., what are the behavioral mechanisms by which they come together? (2) Why do starved L1s aggregate? What selective advantage (if any) do these mechanisms or aggregation itself provide? To aid in answering these questions, we describe here a simple mathematical model for L1 aggregation. In our first report of L1 aggregation behavior \cite{Artyukhin2015}, we speculated on the answers to both question. This paper directly addresses only question (1). 

L1 aggregation is not the only known \cel aggregation behavior, and ours is not the first mathematical model of \cel aggregation. It has been known for many years that in the presence of food (bacteria), most true wild isolates of \cel aggregate, a behavior known as social feeding \citeleft\citen{DeBono1998}\citepunct\citen{DeBono2002}\citeright. Wild strains of \cel prefer low concentrations of oxygen. The usual \cel laboratory strain, N2, does not display social feeding at normal atmospheric oxygen pressure because of a gain-of-function mutation in the neuropeptide receptor gene \textit{npr--1} \cite{DeBono1998}. We and others have speculated that the consumption of oxygen in an aggregate of worms lowers oxygen concentration and thereby attracts more worms \cite{Rogers2006}, although this explanation is disputed \cite{Ding2019}. Mathematical models of social feeding have recently been published \citeleft\citenum{Ding2019}\citepunct\citenum{Demir2020}\citeright. 

A third type of aggregation is mediated by indole-containing ascarosides \cite{VonReuss2012a}. L1s of \textit{daf--22} mutants, which are unable to make ascarosides \citeleft\citenum{Butcher2009}\citepunct\citenum{Golden1985}\citepunct\citenum{Joo2009}\citeright) aggregated similarly to wild type \cite{Artyukhin2015a}. Thus L1 aggregation is different from ascaroside-mediated aggregation. Observations of yet another type of aggregation have recently been published, together with a model \cite{Sugi2019}. This form occurs in the long-term survival form of the worm---the dauer larva---and is probably mediated largely by a simple physical mechanism, surface tension. 

The model we present here is simpler than previous \cel aggregation models in the following sense: it does not describe aggregation behavior in completely realistic detail. We attempt only to reproduce the essential aspects of the behavior. Accordingly, we simply assume the existence, which has been experimentally demonstrated \citeleft\citen{Bargmann2006,Bargmann1997}\citeright, of taxis mechanisms that allow worms to move in the direction they want to go. Although taxis mechanisms have been investigated for years, and much is known about them \citeleft e.g. \citen{Mori1999,Dunn2004,Roberts2016,Pierce-Shimomura1999,Ferree2020}\citeright, the model presented here is based on the idea that the end result of taxis (movement towards favored places) is sufficient to understand aggregation, and that mechanistic details are not essential. 

A further simplification is to describe worms not as individuals, but via population density: a continuous function of space and time, $\rho(t, \vx)$. We also propose a simple mechanism for interactions among worms via diffusible chemical signals. The resulting model takes the form of a system of partial differential equations (PDEs), a variation on the classic Keller-Segel \cite{Keller1970} model, developed to explain the aggregation of cellular slime mold amoebae. Since its original publication, the Keller-Segel model has been the subject of much mathematical analysis \citeleft sections 11.1-11.3 of \citen{Edelstein-Keshet2005} \citepunct\citen{Horstmann2003}\citepunct\citen{Horstmann2004}\citepunct\citen{Hillen2009}\citeright. Indeed, the Keller-Segel model, in its many variations, has become one of the classic models of pattern formation in mathematical biology. This model has the advantage of high mathematical tractability, both analytical \cite{Avery2020} and numerical, the latter of which is the focus of this paper. 

% Results and Discussion can be combined.
\section*{Results}

\subsection*{Strategy}
\label{S:Strategy}

As described in the Introduction, our model is a deliberately simplified description of behavior that assumes the existence of taxis mechanisms that allow worms to move in the direction they want to go. Further, worms are modeled not as individuals, but as a continuous function of space and time, population density $\rho(t, \vx)$. This simplification allows us to describe worm movement with a mathematically tractable partial differential equation (PDE) model similar to the well-studied Keller-Segel \cite{Keller1970} model. Also, because worm density is a component of the model, it is straightforward to implement worm movement that explicitly depends on density. A disadvantage is that individual worms are not accurately represented by a continuous density function. Moreover, we ignore the fact that worms are worm-shaped: i.e., a worm is long and thin, with a head at one end and a tail at the other. Worm geometry is central to some other published models of \cel aggregation \cite{Rogers2006,Ding2019,Sugi2019}. 

We present two versions of the model: the precursor \textit{attractant-only} model and the final \textit{attractant+repellent} model. We begin with the simpler attractant-only model, which is closest in form to the original Keller-Segel model. This model was a partial success---it reproduced certain aspects of L1 aggregation seen in experiments, while failing in others. The more complicated attractant+repellent model better reproduced L1 aggregation behavior.

\subsection*{Design of the PDEs}
\label{S:PDEdesign}

The attractant--only model for L1 aggregation consists of two coupled PDEs, a reaction-diffusion equation that describes the time evolution of the concentration of a diffusible chemical attractant, and a Fokker-Planck equation that describes the movement of the worms. The attractant PDE is:

\begin{equation}
    \label{E:UPDE}
    \dot{U} = U_t = -\gamma U + D\laplacian{U} + s\rho
\end{equation}

\noindent
The worm PDE is:

\begin{align}
    \label{E:wormrhoPDE}
    \dot{\rho} = \rho_t &= \div\left(
        \rho\grad\left(
            V_U(U) + V_\rho(\rho) + \sigma\log\rho
        \right)
    \right)\\
    &= \grad{\rho}\cdot\grad{V} + \rho\laplacian{V} + \sigma\laplacian{\rho}
    \label{E:wormrhoPDE2}
\end{align}

\noindent
Functions and parameters appearing in (\ref{E:UPDE} - \ref{E:wormrhoPDE2}) are listed in Table \ref{T:funcsparams}.

Intutively, the terms of \eqref{E:wormrhoPDE2} can be understood as follows. The first two terms describe how density changes when worms move towards lower potential. The first term, $\grad{\rho}\cdot\grad{V}$ arises from the movement of worms when the density is nonuniform. E.g., if density is low on the left and high on the right ($\grad{\rho} > 0$) and and worms in bulk are moving leftward ($\grad{V} > 0$) the density of worms at any fixed point will increase. The second term $\rho\laplacian{V}$ describes increases in worm density when the worms converge towards a minimum of potential---$\laplacian{V}$ is positive at and near minima. The final term, $\sigma\laplacian{\rho}$ describes changes in density caused by random movement of the worms. Random movement tends to flatten out inequalities, so that density increases near minima of density ($\laplacian{\rho} > 0$) and decreases near maxima ($\laplacian{\rho} < 0$). 

\begin{table}[!ht]
    \caption{\bf Functions and parameters appearing in the PDEs}
    \centering
    \begin{tabular}{l l}
         $\rho(t,\vx)>0$&local density of worms\\
         $U(t,\vx)>0$&local concentration of chemical attractant\\
         $\sigma > 0$&quantifies the random movement of the worms \\
         $\gamma > 0$&spontaneous decay rate of attractant \\
         $D > 0$&diffusion constant of attractant \\
         $s > 0$&is the rate of secretion of attractant per worm\\
         $V_U(U)$&potential function that describes the worm's response to attractant \\
         $V_\rho(\rho)$&potential function that describes the worms' direct response to other worms \\
         $V$&shorthand notation for $V(\rho,U) = V_U(U)+V_\rho(\rho)$ 
    \end{tabular}
    \label{T:funcsparams}
\end{table}

\noindent
These equations are similar to those developed by Keller and Segel \cite{Keller1970} to model the aggregation of \textit{Dictyostelium discoideum} amoebae. Attractant PDE \eqref{E:UPDE} is identical to the reaction-diffusion equation with which they model acrasin. Worm PDE \eqref{E:wormrhoPDE} is a generalization of the equation they use to model the movement of amoebae, which, with specific choices of the potential functions $V_U$ and $V_\rho$, reduces to theirs. 

In designing this model for L1 aggregation, we sought to reproduce certain general characteristics that were obvious in recordings of worm aggregation. First, the worms aggregate. This suggests that they are somehow attracted to each other. Given what we know about \cel biology, it was an obvious guess that this attraction could be mediated by a diffusible chemical signal with limited range \cite{Bargmann1997}. PDE \eqref{E:UPDE} is essentially the simplest physically plausible that meets these criteria. 

The design of the PDE describing the movement of the worms was more complicated. On the time scale of the experiments, neither birth nor death of new worms occurs. This suggested that it should be possible to express the rate of change of worm density as the divergence of some flow field. Since flux occurs by movement of worms, the net flux vector at any point is the density times the mean velocity of worms at that point. These considerations lead to a general equation of the form

\begin{equation}
    \label{E:genrhoPDE1}
    \rho_t = -\div
        \left(\rho\mathbf{v}\right)
\end{equation}

\noindent
in which velocity $\mathbf{v}=\mathbf{v}(\rho,U,\grad{\rho},\grad{U})$ is a vector field depending on density and attractant concentration and their gradients. We chose to assume that the velocity field is conservative, i.e. that it can be represented as the gradient of some scalar potential field. There is no compelling biological necessity for this assumption. We made it for two reasons: First it makes the PDE system more tractable analytically. Second, in recordings of worm behavior, we see that the worms eventually approach an equilibrium in which there is little net flow of animals, and no cyclic flows are obvious. 

If the velocity field $\vv$ is conservative, then it can be expressed as the negative of the gradient of some scalar potential field $V$. We chose a potential function that is a sum of a signal-dependent potential $V_U$ and a density-dependent potential $V_\rho$, for convenience in separately engineering signal and density dependence. This led to the final form \eqref{E:wormrhoPDE}. 

For an attractant, $V_U$ must be a decreasing function of signal. In early simulations with a linear $V_U$ we encountered problems with numerical instability. Steep signal gradients frequently occur in the course of simulation. With a linear $V_U$, these led to large velocities, which meant that worm density at one location was rapidly affected by density at distant locations. As a result it was impractical to satisfy the Courant–Friedrichs–Lewy (CFL) stability condition. We therefore sought a potential function whose dependence on $U$ was convex. 

A modification of Weber's Law that closely corresponds to empirical data is

\begin{equation}
    \label{E:ModWeber}
    \frac{\Delta\phi}{\phi+\alpha} = \beta,
\end{equation}

\noindent
where $\Delta\phi$ is the smallest detectable change in a stimulus $\phi$ and $\alpha$ and $\beta$ are constants,. (This is Eq (1.2) of reference \citen{Gescheider1997}.) If we assume that just noticeable differences in stimulus represent equal changes in sensation $\psi$, we can integrate \eqref{E:ModWeber} to obtain the following psychophysical magnitude function,

\begin{equation}
    \label{E:ModWeberLaw}
    \psi(\phi) \propto \beta \log(\alpha + \phi)
\end{equation}

\noindent
If attractant $U$ is the stimulus and potential $V_U$ the sensation, we get the following potential

\begin{equation}
    \label{E:Vudef}
    V_U(U) \coloneqq -\beta\log(\alpha + U)
\end{equation}

We negate $\beta$ because worms move down a potential, and the potential for an attractant should thus be a decreasing function of its concentration. Parameter $\beta$ determines the strength of attraction. The same potential with negative $\beta$ describes a repellent.

We speculated that the circular shape of the aggregates \cite{Artyukhin2015} results from the worms packing together as tightly as possible. To reproduce this effect in simulations, we designed a density-dependent $V_\rho$ potential that would reflect worms taking up space. The ideal would have been a hard sphere potential

\begin{equation}
    \label{E:Vhsphere}
    V_\rho(\rho) = \left\{\begin{array}{cl}
         0 & \text{if }\rho < \rho_{\text{max}} \\
         \infty & \text{if }\rho \ge \rho_{\text{max}} 
    \end{array}\right.
\end{equation}

\noindent
This potential function implies discontinuous time or spatial derivatives of density, and therefore functions poorly with numerical methods for solving the PDE system. We therefore approximated the discontinuity with a hyperbolic tangent function.

\begin{equation}
    \label{E:Vrhodef2}
    V_\rho(\rho) = \sigma\frac{\mathtt{scale}}{2}\left(
        1 + \tanh\left(
            \frac{\rho - \rho_{\text{max}}}{\mathtt{cushion}}
        \right)
    \right)
\end{equation}

\begin{figure}[!ht]
    \hrule
    \vspace{12pt}
    \begin{center}
    \includegraphics[width=4in]{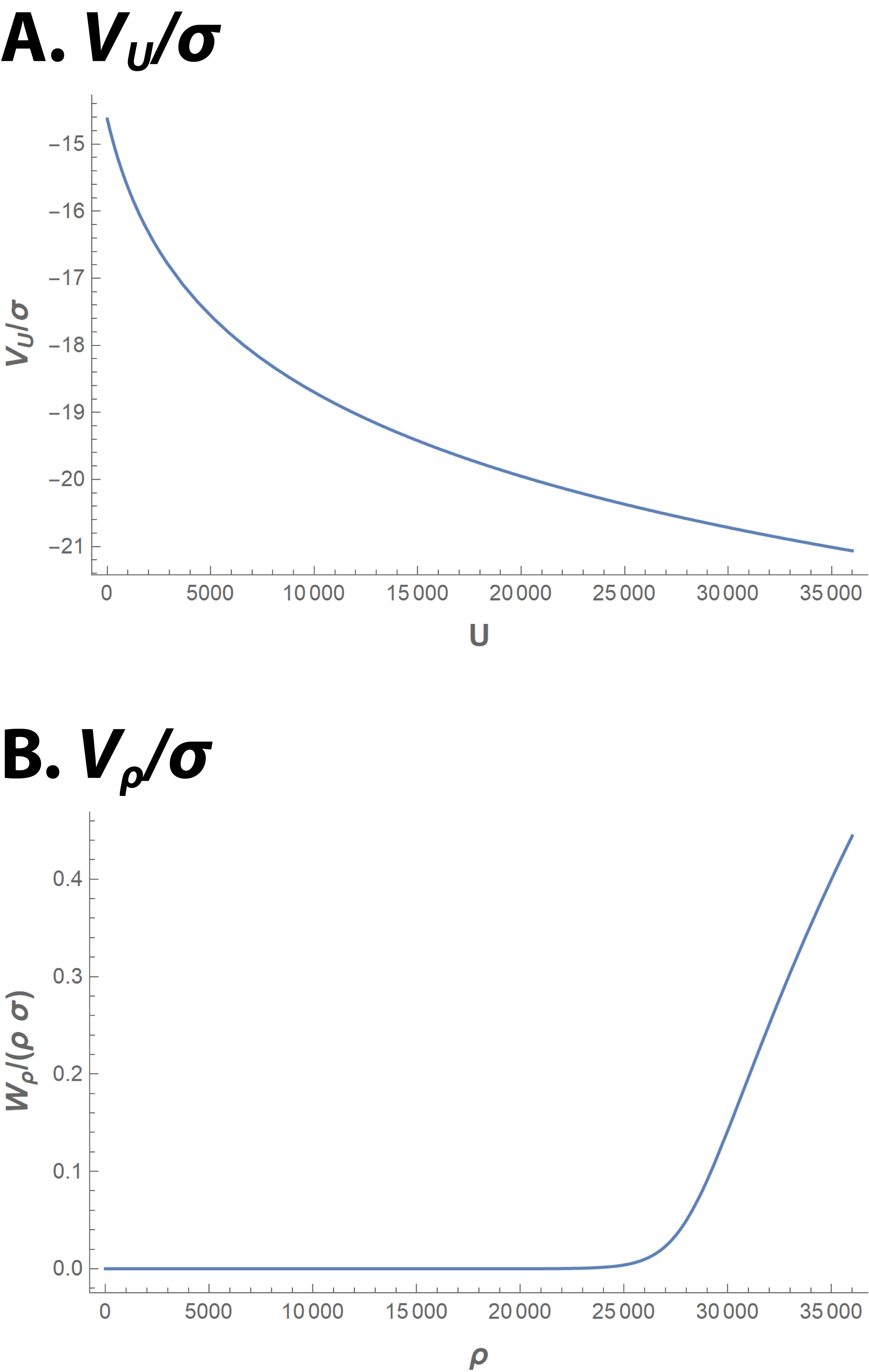}
    \end{center}

    \caption{\bf Potential function plots.}
    Potential functions that appear in the $\rho$ PDE \eqref{E:wormrhoPDE}. Both potentials are made dimensionless by dividing them by $\sigma$. Parameter values are as in Table \ref{T:ch3.params}.
    \label{F:fig3.potentials}
\end{figure}

\noindent
Four parameters determine the exact shape of $V_\rho$: $\sigma$, $\rho_{\text{max}}$, \texttt{scale}, and \texttt{cushion}. (We refer to the latter two by the symbols used to represent them in software code, since they will play little role in the mathematics.) Two parameters, $\sigma$ and \texttt{scale}, determine the vertical scale. $\sigma$ is the parameter that measures random worm movement (see \eqref{E:wormrhoPDE}). $V_\rho$ rises from near 0 for small values of $\rho$ to $\sigma\times\mathtt{scale}$ for large $\rho$. Parameter $\rho_{\text{max}}$ is the density at which $V_\rho$ reaches half its maximum possible value. It is the point at which the $V_\rho$ curve is steepest, and therefore the closest approximation to $\rho_\text{max}$ of \eqref{E:Vhsphere}. Parameter \texttt{cushion} determines how abrupt the rise of $V_\rho$ is. These functions are plotted in Fig \ref{F:fig3.potentials}. The Keller-Segel literature describes other, less-flexible, models in which organisms take up space \cite{Hillen2009}, which we elected not to use. 

\subsection*{Parameter estimates}
\label{S:params}

We required numerical estimates of parameters $\gamma$, $D$, and $s$ that appear in \eqref{E:UPDE} and $\sigma$ of \eqref{E:wormrhoPDE}. In addition, we required values for $\rmax$, \texttt{scale}, \texttt{cushion} and $\alpha$ and $\beta$ which determine the shapes of the potential functions $V_\rho$ and $V_U$. 

A \cel L1 is approximately a cylinder of diameter \SI{15}{\micro\meter} and length \SI{240}{\micro\meter} \cite{Avery2003}. Since worms lie on their sides, a worm occupies approximate area $15\times240 \approx 3600$ \si{\micro\meter^2}. We chose the inverse of this area, \SI{28000}{cm^{-d}} as the parameter $\rmax$. (Here $d = 1\,\text{or}\,2$ is the spatial dimension. The same number, \num{28000}, was used for one and two-dimensional simulations to facilitate comparison.) Parameters \texttt{cushion} and \texttt{scale} have no real biological significance. Parameter \texttt{cushion} makes the ideal hard-sphere potential \eqref{E:Vhsphere} continuous and differentiable, so that the PDEs can be solved numerically with differentiable functions. The value \SI{2000}{cm^{-d}} worked. The \texttt{scale} parameter need only be chosen large enough to constrain the maximum density---we chose 2. 

In small-scale simulations, we chose a mean density $\bar{\rho} = $ \SI{9000}{cm^{-d}} so that aggregates would occupy about $1/3$ (i.e. $9000/28000$) of the area. 

Small molecules in water typically have diffusion constants in the range \SIrange{1e-6}{1e-5}{cm^2 \ s^{-1}}. (We assumed that the signals diffuse through the agar-solidified water under the worms. Worms also respond chemotactically to volatile chemicals diffusing through the air above them \cite{Bargmann1997}---diffusion constants for such volatile signals would be much larger than for water-soluble signals.) We chose the diffusion constant of attractant, $D_a = $\SI{1e-6}{cm^2 \ s^{-1}}, at the lower end of the range of diffusion constants in water. The aggregates that form have diameters of hundreds of micrometers. The mean distance a molecule of attractant diffuses before decaying is $\sqrt{D_a/\gamma_a}$. We therefore chose $\gamma_a$, the decay rate of attractant, to give it a range $\sqrt{D_a/\gamma_a}=$ \SI{100}{\micro\meter}.  To fulfil its role in the model, the repellent, introduced below, needs to have a longer range, so we chose a large diffusion constant of $D_r = $\SI{1e-5}{cm^2 \  s^{-1}} and a smaller decay rate, giving it a range of \SI{1}{mm}. 

Parameters $s_a$ and $s_r$, the rates at which a worm secretes attractant and repellent, effectively set the units of concentration. We chose units of concentration such that $s_i$ and $\gamma_i$ (for $i = a\,\text{ or }\,r$) were numerically equal. (That is, if concentration is measured in ``number of units of stuff''/\si{cm^{d}}, we chose the units in which ``stuff'' is measured to be the amount secreted by one worm in one mean lifetime of the stuff, i.e. $\gamma_i^{-1}$. This has the effect that if $\gamma_i =  \langle \text{number}\,\rangle$ \si{s^{-1}}, then $s_i = \langle \text{number}\rangle\text{``stuff units''}\,$ \si{cm^{-d}s^{-1}}, with the number being the same in the two cases.) This ensures that concentrations $U_i$ and worm density $\rho$ are in the same range numerically.

Artyukhin et al found that the minimum worm density for aggregation is \SI{1500}{cm^{-2}} \cite{Artyukhin2015}. We identified this with the density threshold for instability. We chose $\alpha_a = \alpha_r = $ \SI{1500}{cm^{-2}}, to make $V_U$ linear near the threshold, and to be obviously convex near $\rmax$. We then chose $\beta_a = 2\sigma$ to reproduce the \SI{1500}{cm^{-2}} density threshold for instability in the attractant-only model. For the attractant+repellent model described below, we kept this value for $\beta_a$ and chose $\beta_r = -2\sigma$ for the repellent. Adding repellent to the model increased the calculated instability threshold to \SI{2357}{cm^{-d}}.

Parameter $\sigma$ determines how rapidly the worms spread. Artyukhin et al \cite{Artyukhin2015} found that worms placed at the center of a \SI{6}{cm} diameter petri plate spread to occupy much of the area of the plate in \SI{12}{\hour}, but at this time they still remain mostly concentrated near the center. To estimate $\sigma$, we asked what value of $\sigma$ would reproduce the observed behavior of worms in circular petri plates.

We began by choosing values of $\sigma$ that would approximately reproduce this distribution if the worms' motions were purely diffusional, i.e., if their motions were governed by 

\begin{equation}
    \label{E:diffeq}
    \rho_t = \dot{\rho} = \sigma \laplacian{\rho},
\end{equation}

\noindent
with Neumann boundary condition

\begin{equation}
    \label{E:diskBC}
    \eval{\dv{\rho}{r}}_{r=R} = 0
\end{equation}

\noindent
Here $R =$ \SI{3}{cm} is the radius of the petri plate. Eqs. (\ref{E:diffeq}, \ref{E:diskBC}) can be solved by separation of variables. Any solution $\rho(t, r,\theta)$ can be represented as a sum of exponentially decaying eigenfunctions of the Laplacian. The circularly symmetric eigenfunctions on the disk with Neumann boundary condition \eqref{E:diskBC} are

\begin{equation}
    \label{E:diskef}
    \rho(t, r) = e^{-\sigma k^2 t}J_0(k r),
\end{equation}

\noindent
where $J_0$ is a Bessel function of the first kind. The wavenumber $k$ must be chosen to satisfy boundary condition \eqref{E:diskBC}, i.e. $k = j_{1,n}/R$, where $j_{1,n}$ is the $n$\textsuperscript{th} nontrivial zero of $J_1$. The smallest wavenumber, corresponding to the circularly symmetric eigenfunction that decays most slowly, is thus $k = j_{1,1}/R\,\approx\,3.8317/$\SI{3}{cm}$\,\approx\,$\SI{1.28}{cm^{-1}}. We began by choosing $\sigma$ so that the corresponding time constant $\sigma^{-1} j_{1,1}^{-2}R^2$ was approximately \SI{12}{\hour}. Of course, the motion of the worms is not purely diffusional. We therefore refined our estimate of $\sigma$ by numerical solution of the attractant+repellent system described below, from an initial condition in which the worms began near the center of the petri plate. From these simulations we chose $\sigma = $\SI{5.555e-6}{cm^2 s^{-1}} as producing results that resembled experimental results.

Table \ref{T:ch3.params} summarizes parameter values.

\begin{table}
\hrule
\vspace{6pt}
\centering
    \caption{\textbf{Parameter values}}
    \label{T:ch3.params}
\begin{tabular}{l l l}
     $\bar{\rho}$
     &mean worm density
     &\SI{9000}{\centi\meter^{-d}}\\
     $\sigma$
     &random worm movement
     &\SI{5.555e-6}{cm^2 s^{-1}} \\
     $\rmax$
     &midpoint of $V_\rho$ potential rise
     &\SI{28000}{cm^{-d}} \\
     \texttt{cushion}
     &breadth of $V_\rho$ rise
     &\SI{2000}{cm^{-d}} \\
     \texttt{scale}
     &height of $V_\rho$ rise
     &2 \\
     $\beta_a$
     &strength of attraction
     &\SI{1.111e-5}{cm^2 s^{-1}} \\
     $\alpha_a$
     &attractant concentration scale
     &\SI{1500}{cm^{-d}} \\
     $\gamma_a$
     &attractant decay rate
     &\SI{0.01}{s^{-1}} \\
     $D_a$
     &attractant diffusion constant
     &\SI{1e-6}{cm^2 s^{-1}} \\
     $s_a$
     &attractant secretion rate
     &\SI{0.01}{cm^{-d} s^{-1}} \\
     $\beta_r$
     &strength of repulsion
     &\SI{-1.111e-5}{cm^2 s^{-1}} \\
     $\alpha_r$
     &repellent concentration scale
     &\SI{1500}{cm^{-d}} \\
     $\gamma_r$
     &repellent decay rate
     &\SI{0.001}{s^{-1}} \\
     $D_r$
     &repellent diffusion constant
     &\SI{1e-5}{cm^2 s^{-1}} \\
     $s_a$
     &repellent secretion rate
     &\SI{0.001}{cm^{-d} s^{-1}}
\end{tabular}
    \hrule
    \vspace{6pt}
\end{table}

\subsection*{Simulation results, attractant-only model}

\begin{figure}
    \hrule
    \vspace{12pt}
    \begin{center}
    \includegraphics[width=5in]{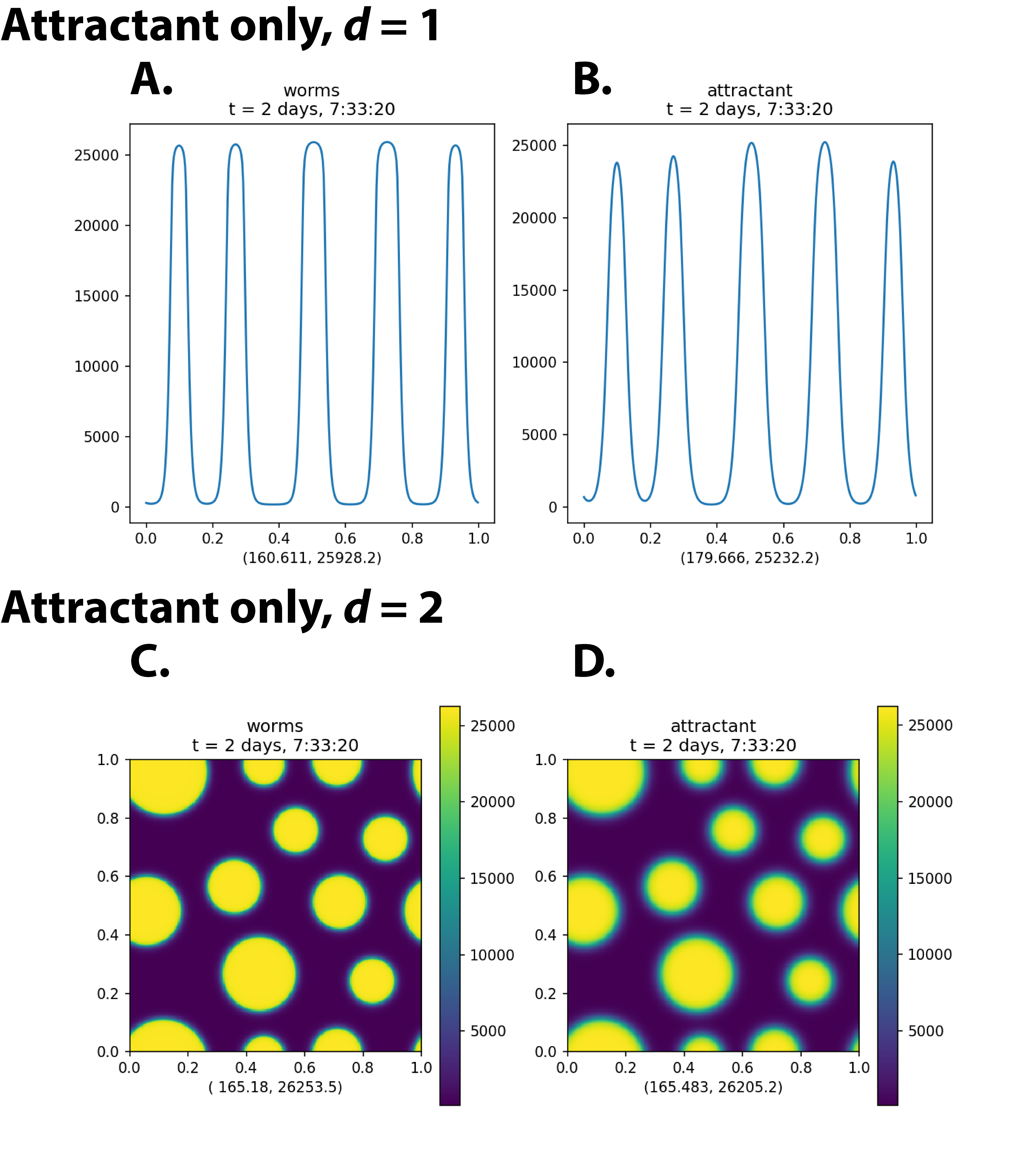}
    \end{center}
    
    \caption{\textbf{Simulation of the attractant-only model}}
    This figure shows the state of a numerical simulation of the attractant-only model after \SI{200000}{s} (2 days and 7 hours). The initial condition was a uniform worm density of $\bar{\rho} = $ \SI{9000}{cm^{-d}}, perturbed by normally distributed random noise of standard deviation 1\% (i.e. \SI{90}{cm^{-d}}). (The entire time courses can be seen in the videos \texttt{options138a.mp4} and \texttt{options139.mp4} in the Supporting Information.) Supporting Information Fig \ref{F:fig3.AttractantReruns} shows results at $t = $ \SI{200000}{s} and $t = $ \SI{1e7}{s} (116 days) of ten independent runs of the same simulation with different pseudorandom noise in the initial condition.
    Panels \textbf{A, B} show the results of simulations in one-dimensional space; \textbf{C, D} show results in two-dimensional space. \textbf{A, C} show density $\rho$; \textbf{B, D} show attractant concentration $U$. The two numbers below each plot are the minimum and maximum values of the plotted function over the entire \SI{1 x 1}{cm} domain. The spatial units are centimeters.
    \label{F:fig3.periodic}
    \vspace{6pt}
    \hrule
\end{figure}

Fig \ref{F:fig3.periodic} shows results of numerical simulations of the attractant-only model.

%1d: options138a
%2d: options139a
%AttractOnly.ai,.png

This model successfully reproduced the experimental results in certain ways, but failed in others. It was successful in that circular aggregates of maximum density rapidly formed. The aggregates had sharp boundaries, and outside of aggregates worm density was low and uniform. This is most easily seen in the one-dimensional results (Fig \ref{F:fig3.periodic}A), but is also true in two dimensions. These are also characteristics of the experimental results. (See Supporting Information video \texttt{N2\_5e5\_washed.avi}.)

The model failed to reproduce the patterning of aggregates. In experiments (Supporting Information video \texttt{N2\_5e5\_washed.avi}), aggregation appears to reach an equilibrium after \SI{12}{h}. Although individual worms continue to move actively, the aggregates themselves show little change after the first several hours. Those aggregates that form after the worms disperse from where they are initially placed are never larger than ca. \SI{700}{\micro m} in diameter. Most of the worms, in fact, end up in aggregates close to this maximum size. These aggregates are also fairly uniformly spaced---the distance from one aggregate to its nearest neighbors varies little. 

In numerical solutions of the attractant-only model, however, aggregates had no maximum size (aside from that imposed by the fixed finite number of worms), and their spacing was not uniform. Furthermore, even after \SI{200000}{s}, they were not at equilibrium. This can be seen by computing the velocity $v = \norm{\grad{V}}$, which at equilibrium would be zero everywhere, but remained well above zero throughout the simulation. More obviously, it is seen by continuing the solution past $t = $ \SI{200000}{s}. Fig \ref{F:fig3.AttractantReruns} shows that aggregates increased in size and decreased in number between $t = $ \SI{200000}{s} and $t = $ \SI{1e7}{s}. In fact, we believe the only true equilibria of the attractant-only model are those in which there is a single large aggregate containing almost all the worms. This state was reached at $t = $ \SI{1e7}{s} in one of the ten simulations in Fig \ref{F:fig3.AttractantReruns}. 

In fact, this observation is consistent with linear stability analysis of the attractant-only model (see Supplemental Information). PDE system (\ref{E:UPDE}, \ref{E:wormrhoPDE}) shares with the original Keller-Segel system the property of density-dependent instability. The condition for a sinusoidal variation of wavenumber $k$ to be unstable is \eqref{E:instab1}. The condition for instability \eqref{E:instab1} has no minimum wavenumber other than zero. Wavenumber is inversely proportional to wavelength, so that there is no nonzero minimum wavenumber means there is no natural maximum size for the aggregates that form when the density exceeds threshold. (This is a well-known property of the classical Keller-Segel model as well---see, for instance, section 11.3 of reference \citen{Edelstein-Keshet2005}.) It is true that the attractant has a natural range, $\sqrt{D/\gamma}$---the distance an average molecule diffuses before it decays. However, worms attract each other, albeit weakly, even when they are further apart than this. There is thus no mechanism in the model (\ref{E:UPDE}, \ref{E:wormrhoPDE}) that would prevent the merging of aggregates to unlimited size. This is true for any attractant-only Keller-Segel model.

\subsection*{A repellent is necessary}

We could not reproduce the experimental observed uniformity of aggregate size in numerical experiments with attractant-only models. We suspected that the addition of a negative signal to oppose the attractant, a repellent, would solve the scale problem. Linear stability analysis supports this intuition. (See Supporting information section Linear stability analysis of the attractant+repellent model.)

Intuitively, what one requires is a short-range attractant and a long-range repellent. We therefore added to the attractant-only model a repellent with diffusion constant $D_r = 10D_a$ and decay rate $\gamma_r = 0.1\gamma_a$. The range of this repellent $\kappa_r=\sqrt{D_r/\gamma_r} =$ \SI{1}{mm} is ten times that of attractant, and is approximately equal to the observed spacing between aggregates. Thus, in the attractant+repellent model, attractant PDE \eqref{E:UPDE} is replaced with two PDEs, one \eqref{E:UaPDE} for attractant and the other \eqref{E:UrPDE} for repellent. 

\begin{align}
    \label{E:UaPDE}
    \dot{U_a} &= -\gamma_a U_a + D_a\laplacian{U_a} + s_a\rho \\ 
    \dot{U_r} &= -\gamma_r U_r + D_r\laplacian{U_r} + s_r\rho
    \label{E:UrPDE}
\end{align}

%1d: options140a
%2d: options141a
%AttractantRepellent.ai,.png

\begin{figure}[!ht]
    \hrule
    \vspace{12pt}
    \begin{center}
    \includegraphics[width=5in]{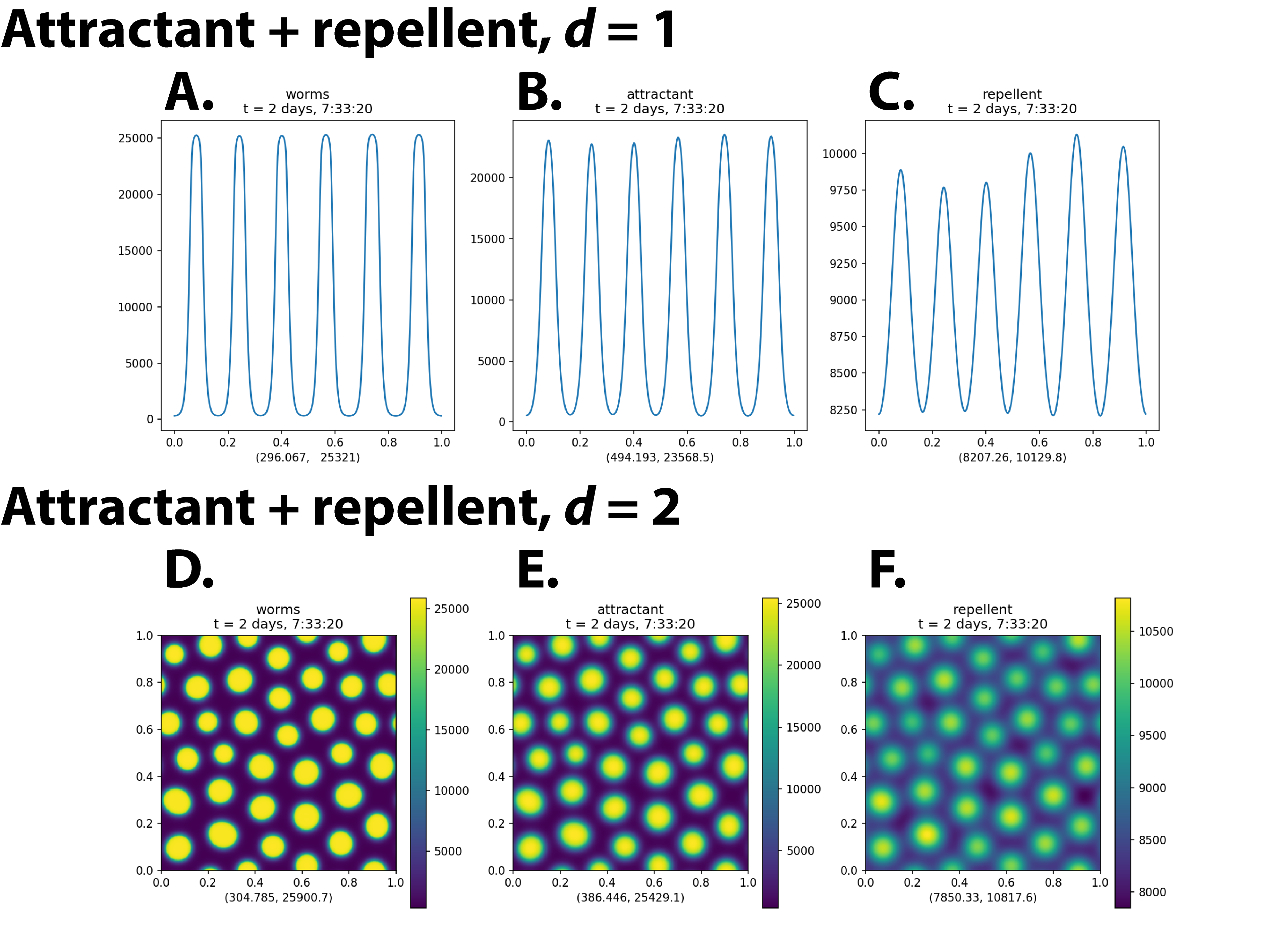}
    \end{center}

    \hrule
    \vspace{6pt}

    \caption{Attractant+repellent simulation}
    Panels \textbf{A, D} show density $\rho$, \textbf{B, E} show attractant concentration $U_a$, and \textbf{C, F} show repellent concentration $U_r$. The spatial units are centimeters. The two numbers below each plot are the minimum and maximum values of the plotted function over the entire \SI{1 x 1}{cm} domain. Note the different scale of the attractant and repellent plots. The means are the same, but because repellent is a longer-range signal, it is smoothed much more by diffusion and varies less than attractant. 
    (Supporting Information videos \texttt{options140a.mp4} and \texttt{options141.mp4} show the full time courses for these simulations. Supporting Information Fig \ref{F:fig3.AttractantRepellentReruns} shows ten independent solutions of the two-dimensional system with different pseudorandom noise at time 0.) 
    \label{F:fig3.repellent}
    \vspace{6pt}
    \hrule
\end{figure}

As shown in Fig \ref{F:fig3.repellent}, addition of a repellent to the model produced the predicted effect. Aggregates formed with characteristic size and spacing approximately matching those seen in experiments on worms. These solutions are close to equilibrium at $t = $ \SI{200000}{s}, as seen by comparison with the results at $t = $ \SI{1e7}{s}. (Compare Figs \ref{F:fig3.AttractantRepellentReruns}A and B.)
There is even a hint of pattern formation, with the aggregates in an approximate hexagonal array at $t = $ \SI{200000}{s}. The hexagonal patterning is near perfect at $t = $ \SI{1e7}{s}, with the exception of lattice defects, most easily recognized as slightly smaller aggregates surrounded by five rather than six neighbors. (Fourier analysis confirms the regularity of these patterns \cite{Avery2020}.) If there is a problem with the result, it is that the array is too perfect. More irregularity is seen in experiments with real animals.

\subsection*{\srhtwo encodes a G--protein coupled receptor expressed in starving L1s}

\begin{figure}[!ht]
    \hrule
    \vspace{12pt}
    \begin{center}
    \includegraphics[width=5in]{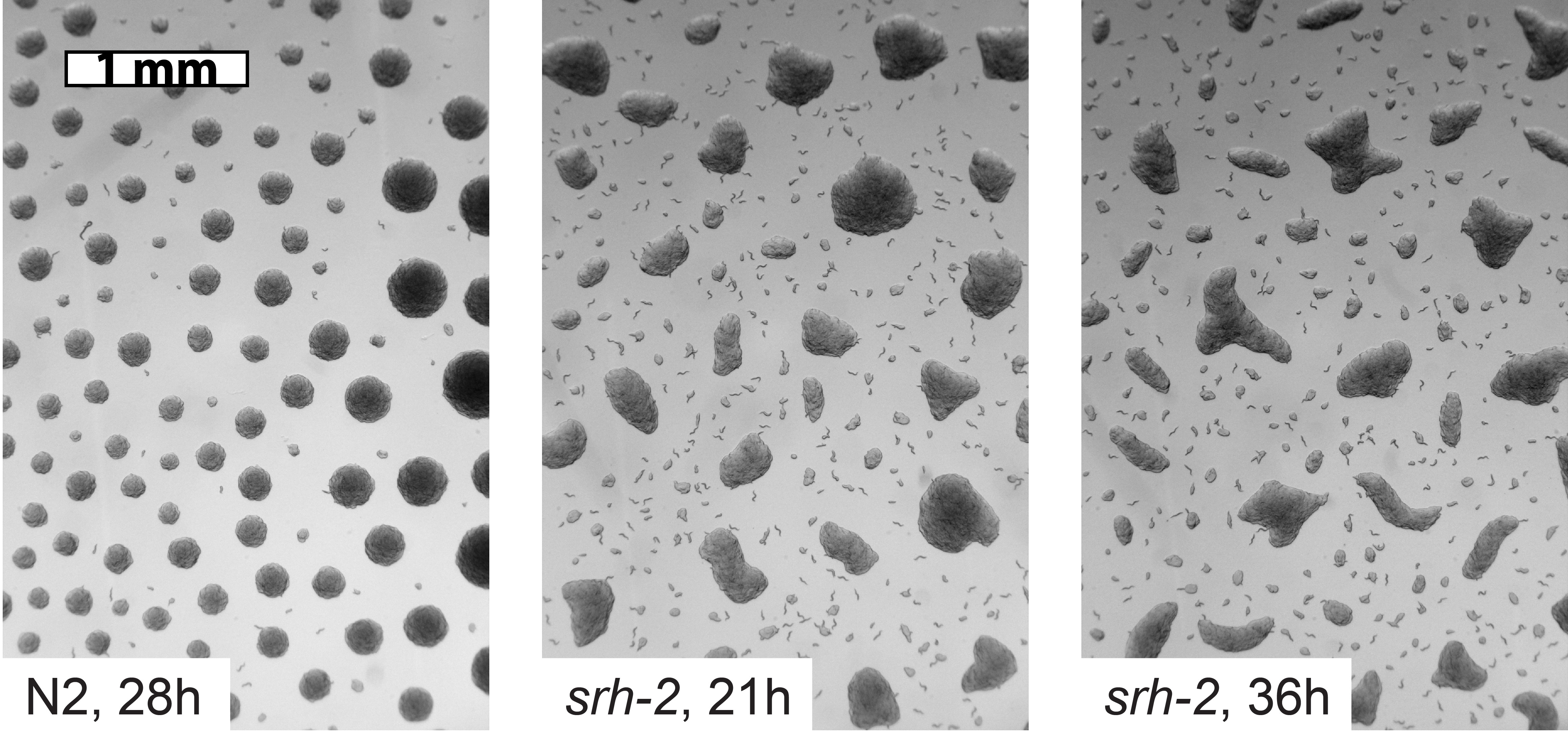}
    \end{center}

    \hrule
    \vspace{6pt}

    \caption{\srhtwo knockout L1 aggregation}
    Starved L1s of mutant worms lacking a functional \srhtwo gene aggregate, but the aggregates they form are irregularly shaped (the \textit{animal crackers} phenotype).
    \label{F:srh-2}
    \vspace{6pt}
    \hrule
\end{figure}

Attempting to understand molecular mechanisms of L1 aggregation, we measured gene expression in starved L1s in the presence and absence of ethanol or acetate, either of which is required for aggregation \cite{Artyukhin2015}. We identified an ethanol-induced gene, \srhtwo, whose expression increases at the time that starved L1s become capable of aggregation (Supporting information). Gene \srhtwo is predicted to encode a sensory receptor, i.e., a protein expressed on the surface of a sensory neuron, capable of detecting chemicals in the environment. To find out whether \srhtwo plays a role in L1 aggregation, we knocked the gene out. (That is, we genetically engineered a mutant strain that lacks a functional \srhtwo gene.) We then tested the \srhtwo knockout worms for aggregation. As shown in Fig \ref{F:srh-2}, these mutant worms
still aggregate, but the aggregates are irregular in shape. Furthermore, the number of worms outside large aggregates is larger in \srhtwo than in wild-type. In Fig \ref{F:srh-2}B,C one can see that the frequency of isolated individuals and of small aggregates are both elevated. 

\subsection*{The Srh--2 phenotype may be modeled by rapid decay of worm movement}

Two observations suggested a partial explanation of the \srhtwo phenotype. First, in the movies of the attractant+repellent simulation, one sees formation of irregularly shaped aggregates at early times. With time, these aggregates become circular. Second, in movies of aggregating L1s, there is a lot of rapid movement at early times, but as time goes on, fewer worms are seen moving. This suggested that worm movement might slow with time, perhaps because the worms run low on energy. (They are, after all, starving.) 

Together, these observations suggested an explanation for the Srh--2 phenotype--—perhaps the movement of \srhtwo knockout worms slows down faster. In the model, such a movement slowdown would be reflected in the decrease of the parameters $\sigma$ (representing random worm movement) and $\beta_a, \beta_r$ (representing signal--responsive movement) with time. We modeled slowdown with the attractant+repellent PDEs (\ref{E:wormrhoPDE}, \ref{E:UaPDE}, \ref{E:UrPDE}), but with parameters $\sigma, \:\beta_a, \:\beta_r$ time-dependent:

\begin{align}
    \label{E:sigmacool}
    \sigma &= \sigma(t) = \sigma_0 e^{-t/\tau} \\
    \beta_a &= \beta_a(t) = \beta_{a,0} e^{-t/\tau} \\
    \beta_r &= \beta_r(t) = \beta_{r,0} e^{-t/\tau}
\end{align}

\begin{figure}[!ht]
    \hrule
    \vspace{12pt}
    \begin{center}
    \includegraphics[width=5in]{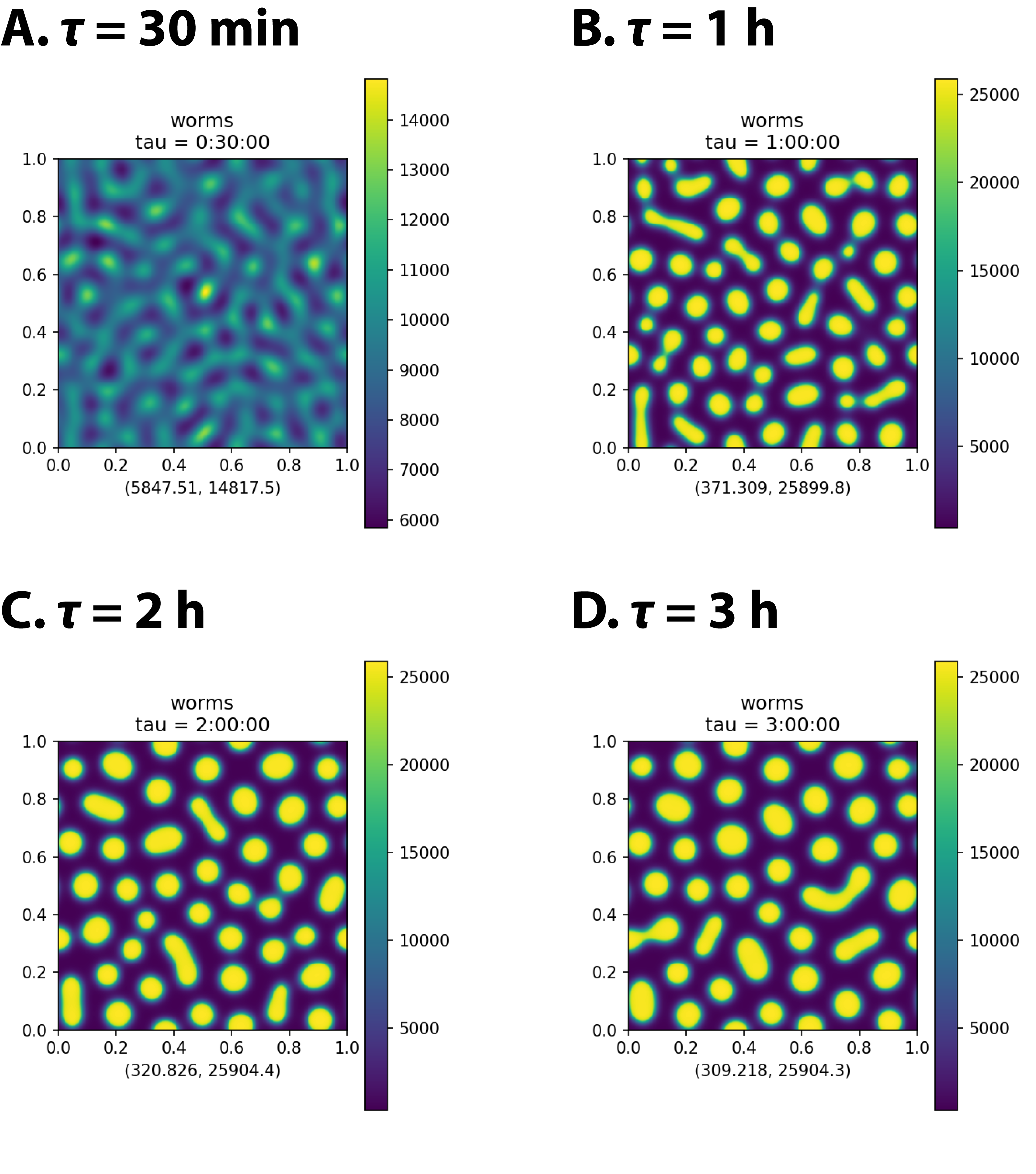}
    \end{center}

    \hrule
    \vspace{6pt}

    \caption{Attractant+repellent simulation with slowdown}
    Worm density $\rho(t, \vx)$ of a slowdown model simulation at $t = $ \SI{200000}{s} for four different values of $\tau$. 
    \label{F:cooldown}
    \vspace{6pt}
    \hrule
\end{figure}

\noindent
(Note that this is not the same as simply stretching the time axis of the attractant+repellent model, because the time--scales of Eqs (\ref{E:UaPDE}, \ref{E:UrPDE}) remain unchanged.)
The $t = 0$ values $\sigma(0), \:\beta_a(0), \:\beta_r(0)$ were the same as those of $\sigma, \:\beta_a, \:\beta_r$ in Table \ref{T:ch3.params}. Fig \ref{F:cooldown} shows results at $t = $ \SI{200000}{s} of simulations of the slowdown model with four different values of $\tau$. When $\tau$ is very small (e.g. \SI{30}{min}, Fig \ref{F:cooldown}A) aggregation is arrested before dense aggregates form. Larger values of $\tau$ permit the formation and persistence of irregular aggregates. For small enough values of $\tau$ (Fig \ref{F:cooldown}A,B), we also see an elevated background worm density. 

\subsection*{Individual-based simulations}

To check our PDE model, we also simulated a cellular Potts individual-based model in the Morpheus \cite{Starruss2014} modeling environment  (details in Methods section). Fig \ref{F:fig3.CPM} shows result that can be compared to Figs \ref{F:fig3.periodic}C,D and \ref{F:fig3.repellent}D,E,F. The results are similar. It has not yet been computationally feasible to reproduce Figs \ref{F:fig3.AttractantReruns}, \ref{F:fig3.AttractantRepellentReruns}, and \ref{F:fullscale}. (These computations are in progress.) 

\subsection*{Full-scale simulations}

\begin{figure}[!ht]
    \hrule
    \vspace{12pt}
    \begin{center}
    \includegraphics[width=5in]{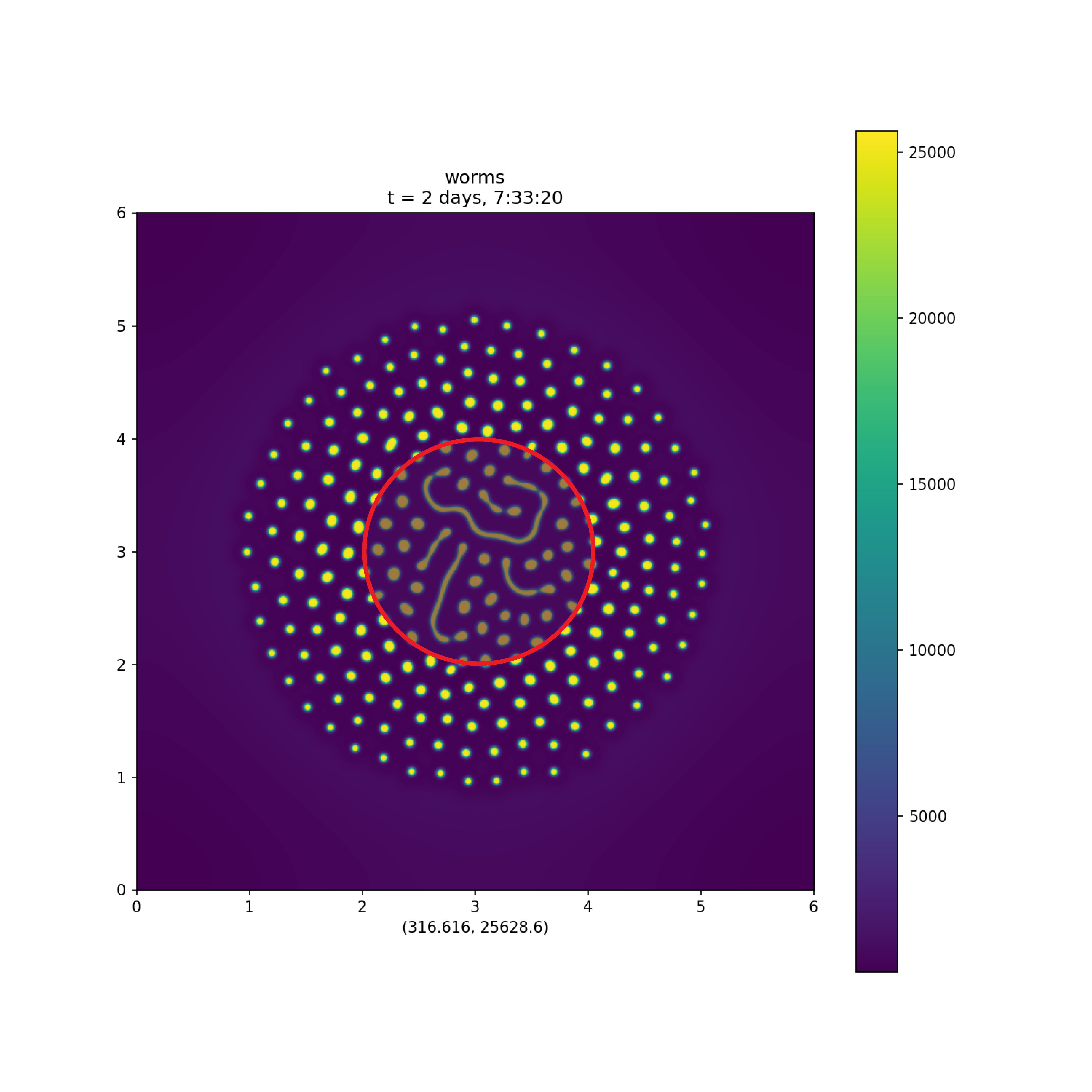}
    \end{center}

    \hrule
    \vspace{6pt}

    \caption{Full-scale attractant+repellent simulation}
    Simulation of the attractant+repellent model on a \SI{6 x 6}{cm} domain. The simulation began with \num{68400}  worms in a \SI{2}{cm} diameter circle at the center of the plate (inner red circle). (Supporting Information video \texttt{options157.mp4} shows the entire time course.)
    The spatial units are centimeters. The two numbers below the plot are the minimum and maximum values of the plotted function over the entire \SI{6 x 6}{cm} domain. Density is muted in a central \SI{2}{cm} diameter circle, corresponding to where the worms were initially placed, to suggest the region in which we think influences not included in our model might be important.
    \label{F:fullscale}
    \vspace{6pt}
    \hrule
\end{figure}

Our experimental studies of aggregation usually begin by placing a large number of worms in the center of a \SI{6}{cm} petri plate \cite{Artyukhin2015a}. (See Supporting video \texttt{N2\_5e5\_washed.avi} which records the behavior over \SI{12}{\hour} of \num{500000} worms that were placed on the center of a plate at time 0.) These experiments begin with the dispersal of the worms, so that the density is high near the center of the plate and lower towards the edges. To more closely mimic such experiments, we solved the attractant+repellent model on a \SI{6 x 6}{cm} square, from an initial condition in which most of the worms began near the center of the plate. Fig \ref{F:fullscale} shows the distribution of worms at $t = $ \SI{200000}{s}. 

We do not believe that the attractant+repellent model accurately represents the physics and biology of worm motions in the region near the center of the plate. For instance, in the preparation of eggs from which the L1s used for the experiment hatch, some non-living debris is inevitably generated. This debris, which is transferred to the center of the plate along with the worms, may influence behavior.

Outside this central region, the behavior of the full-scale simulation resembled the behavior of worms on a \SI{6}{cm} diameter petri plate.

\subsection*{Spectral comparison of experimental and simulation results}

\begin{figure}[!ht]
    \hrule
    \vspace{12pt}
    \begin{center}
    \includegraphics[width=5in]{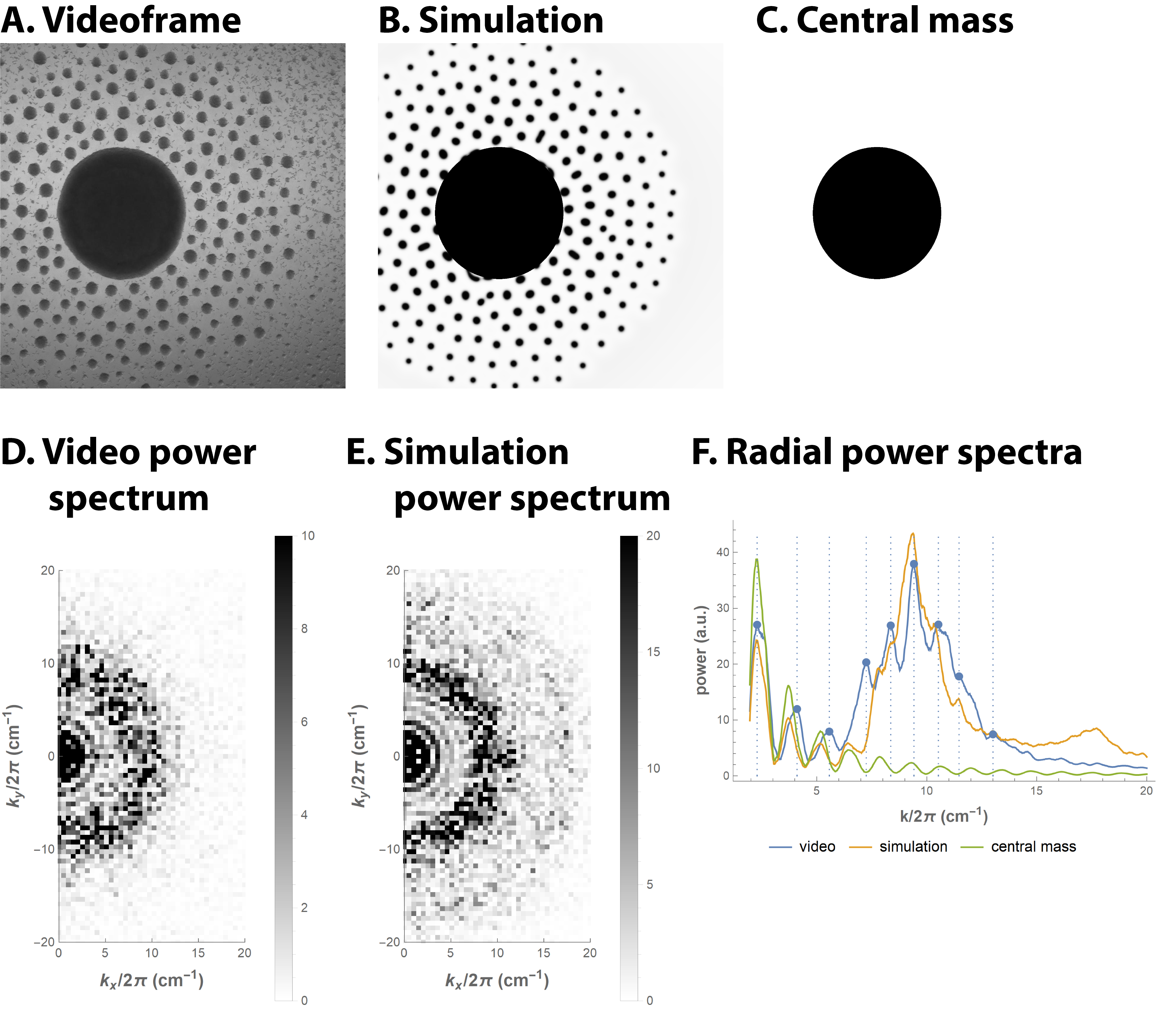}
    \end{center}

    \hrule
    \vspace{6pt}

    \caption{Spectral comparison of experimental and simulation results}
    \textbf{A.} Last frame of aggregation video N2\_5e5\_washed.avi, cropped to a square to facilitate Fourier analysis. This square is \SI{1.93}{cm} in size. \textbf{B.} Final time point of the full-scale simulation \ref{F:fullscale}, cropped and scaled to match \textbf{A} as closely as possible, with a corresponding central mass added. \textbf{C.} The central mass alone. \textbf{D.} Low-frequency region of the two-dimensional power spectrum of the discrete Fourier transform of image \textbf{A}. The power scale is truncated at 10 AU (arbitrary units) \textbf{E.} Low-frequency region of the power spectrum of image \textbf{B}. The power scale is truncated at 20 AU. \textbf{F.} Radially summed power spectra of images \textbf{A},\textbf{B}, and \textbf{C}. Peaks of the experimental image spectrum are picked out in blue.
    \label{F:spectralAnalysis}
    \vspace{6pt}
    \hrule
\end{figure}

As a first approach to quantitative evaluation of the similarity of simulation results to experimental, we compared Fourier Power spectra of the final image of experimental video \texttt{N2\_5e5\_washed.avi} to the \SI{200000}{s} density function from full-scale simulation (Fig \ref{F:spectralAnalysis}). (See the Methods section below for details.)  Two-dimensional spectra (Fig \ref{F:spectralAnalysis}C,D) show prominent rings at wavenumber $k = \norm{\vk} \approx$ \SI{10}{cm^{-1}}, showing the existence of periodic structure with wavelength approximately \SI{1}{mm}. The approximate circular symmetry of the power spectra result from the approximate circular symmetry of the images. 

To obtain higher resolution spectra, we computed radial power spectra by summing power $p_{\vk}$ for spectral components with equal or approximately equal $k = \norm{\vk}$. These radial spectra (Fig \ref{F:spectralAnalysis}F) show additional substructure within the $k \approx $ \SI{10}{cm^{-1}} ring. The radial spectrum of the experimental image has about six peaks between 7 and 13 \si{cm^{-1}}. The simulation spectrum has peaks or shoulders at the locations of some but not all of the experimental spectrum peaks. (Note that we adjusted the scaling of the simulation image to make the largest peaks near \SI{9.5}{cm^{-1}} match, so the coincidence of this particular peak is not significant.) 

In both Fig \ref{F:spectralAnalysis}E and F a weak feature is visible in simulation results at about \SI{18}{cm^{-1}}. Rather than suggesting structures with wavelength \SI{0.5}{mm}, this may be a harmonic of the \SI{1}{mm} periodicity. No such feature is visible in experimental results. It may be that the experimental data are too noisy for such small structures to survive in the power spectrum. 

\section*{Discussion}

\subsection*{Summary}

Our final model, the attractant+repellent model, appeared to reproduce the main features of L1 aggregation. Spectral analysis suggests that the spatial patterning of aggregates in simulations resembles experimental results, although further work along these lines will be necessary. This model is minimal, we believe, in the sense that no simpler model of the Keller-Segel form can adequately reproduce the experimentally observed behavior. In particular, the observed patterning of aggregates---their roughly uniform spacing and sizes, required a short-range attractive influence and a longer-range repulsive influence. Luca et. al. \cite{Luca2003} similarly conclude that a two-signal model is necessary to reproduce the patterning of senile plaques in their Keller-Segel model of Alzheimer disease. 

In our model the attractive and repulsive influences took the form of labile diffusible small molecules produced by the worms. Some other possibilities that can be imagined work equally well. For instance, the attractant could be replaced by a ubiquitous repellent that is locally destroyed by the worms. (In some explanations of social feeding, oxygen plays this role.) It is even possible that the attractant and repellent are the same molecule, if it has the unusual behavioral characteristic of being repulsive at long range (i.e., lower concentration) and attractive at short range (high concentration). Either the attractant or the repellent could be replaced with a physical force, e.g. the physical attraction produced by surface tension (which, however, has too short a range to work well in our current models). 

We also report here new experimental results: the possible sensory receptor gene \srhtwo is expressed under conditions where L1 aggregation takes place. Mutant worms whose \srhtwo gene has been knocked out aggregate, but their aggregates are irregularly shaped, unlike the uniformly circular aggregates of wild-type worms. Our modeling suggests this phenotype could be explained by a faster-than-normal decrease in worm movement in the mutant. This observation is potentially testable by tracking the movement of individual fluorescently labeled worms. 

\subsection*{Validity of the continuum approximation}

Two useful approaches to modeling the movements of populations are individual-based models and continuum approximations. In the individual-based model (also known as a Lagrangian model by analogy to classical mechanics) the population is represented as a collection of agents, each of which moves and changes state according to its own biological imperatives. In the continuum approximation (also known as a Eulerian model) the population is instead represented as a continuous density function of state variables, such as position. As the population is in fact composed of individuals, the individual-based model has greater \textit{prima facie} validity. Continuum models, however, are often more tractable numerically and analytically. 

We used both approaches but relied most heavily on continuum models. How valid is the continuum approximation in this case? To what extent does it distort our results? The clearest way in which continuum results (Fig \ref{F:fig3.periodic}, \ref{F:fig3.repellent}) differ from experimental results (Fig \ref{F:spectralAnalysis}A) and an individual-based model (Fig \ref{F:fig3.CPM}) are the direct effect of the continuum approximation: the density function $\rho(\vx)$, which is constrained to be continuous and differentiable, is smooth, while the actual distribution of worms is inevitably lumpy. This shows up in two obvious ways. First, aggregates in continuum simulations have smooth edges. The edges of the aggregates in experiments have worm-shaped irregularities. Similarly, the aggregates in individual-based simulations show irregularities shaped like simulated worms. 

Second, regions of low density in individual-based simulation or experimental results are mostly empty, but with entire worms dotted here and there. Continuum simulations, in contrast, show regions of uniform low $\rho$. In these regions, however, the continuum results are not so much wrong as subject to a more subtle interpretation. That is $\rho(\vx)$ is best understood as a measure. That is, for a region $R\subset\mathbb{R}^d$, let

\begin{equation}
    \label{E:Enumber}
    \langle N(R) \rangle = \int_R \rho(\vx)\dd\vx
\end{equation}

Then $\langle N(R)\rangle$ is the expected number of worms in region $R$. Where $\langle N(R) \rangle \ll 1$, it can be understood as the probability of finding one worm in region $R$.

Mogilner et al \cite{Mogilner2003}, in studying an individual-based Keller-Segel model state the following criterion, ``A requirement for the validity of the Eulerian approximation is that many organisms are located on a spatial scale on the order of the range of interactions.'' I.e., for the continuum approximation to be valid, no individual worm should matter very much. The continuum approximation is valid at most to the extent that, if we were to remove a single worm, none of the other worms would notice. 

In our models the range of attractant is $r_a = \sqrt{D_a/\gamma_a} = $ \SI{0.01}{cm} $=$ \SI{100}{\micro\meter}. At density $\rho$, a crude estimate of the sphere of influence of a worm---the number of worms in two-dimensional space influenced by a single worm, as well as the number of worms that influence a single worm---is the number within a circle of radius $r_a = $ \SI{0.01}{cm}, $\rho\pi r^2$. Since the mean density in the full-scale simulation (Fig \ref{F:fullscale}) is \SI{2000}{cm^{-2}}, this suggests that the sphere of influence of one worm is \num{0.63} worms, which is certainly not ``many organisms''. This calculation, however, is obviously misleading. A glance at Fig \ref{F:fullscale} shows that the mean density is low because much of the surface is empty space. Mean density $\bar{\rho} = $ \SI{2000}{cm^{-2}} is the density weighted by surface area, i.e., it is an average in which every square millimeter of real estate counts equally. Since we are interested in the worms rather than the agar surface, we should instead compute a worm-weighted average,

\begin{equation}
    \label{E:wwrho}
    \bar{\rho}_w \coloneqq \frac{1}{N}\sum_{i=1}^{N}\rho(\vx_i),
\end{equation}

\noindent
where $N$ is the total number of worms and $\vx_i$ is the location of worm $i$. The number of worms within the sphere of influence of an average worm in Fig \ref{F:fullscale} is then $\bar{\rho}_w\pi r_a^2 = $ \num{4.1}. 

Mogilner et al's \cite{Mogilner2003} analysis is not directly applicable to our model because they make the analytically convenient but psychophysically implausible assumption that influences are additive. This assumption doesn't hold here because potential \eqref{E:Vudef} is nonlinear. To apply their criterion we have to linearize potential around the attractant concentration the worms experience. At the worm-weighted mean attractant concentration (defined analogously to \eqref{E:wwrho}) $\overline{U_a}_w \approx$ \SI{11600}{cm^{-2}}, the elasticity of $V_{U_a}$ is

\begin{equation}
    \label{E:Uelasticity}
    E_{V_{U_a}} \coloneqq \frac{U_a}{V_{U_a}\qty(\overline{U_a}_w)-V_{U_a}(0)}\dv{V_{U_a}}{U_a}\qty(\overline{U_a}_w) \approx 0.41
\end{equation}

\noindent
Thus, if you remove one worm, the worms in its vicinity feel a relative change in attractant potential equivalent to about $0.41/4.1 = 10\%$ of the total attractant effect. These calculations suggest that the continuum approximation may not be drastically inaccurate.

%In[265]:= \[Rho]weightedMeans/@{d138sm1,d139sm1,d140sm1,d141sm1,d157sm1}
%Out[265]= {{22549.5,21166.3},{23385.2,22272.5},{20090.6,18269.7,9649.18},{20287.2,18420.1,9653.03},{12939.7,11639.4,4129.34}}

%In[269]:= \[Rho]weightedMedians/@{d138sm1,d139sm1,d140sm1,d141sm1,d157sm1}
%Out[269]= {{25595.4,23550.4},{26035.,25034.3},{24353.2,20763.4,9782.88},{24849.4,20654.2,9691.72},{13604.2,12840.4,4526.97}}

%In[273]:= countElasticity/@(\[Rho]weightedMeans/@{d138sm1,d139sm1,d140sm1,d141sm1,d157sm1})//Column
%Out[273]= {{7.08414,0.343897,20.5996}}
%{{7.34667,0.339081,21.6664}}
%{{6.31165,0.358371,17.6121},{631.165,0.431458,1462.87}}
%{{6.37342,0.357542,17.8256},{637.342,0.431406,1477.36}}
%{{4.06514,0.408193,9.95887},{406.514,0.554649,732.921}}

%In[282]:= countElasticity/@(\[Rho]weightedMedians/@{d138sm1,d139sm1,d140sm1,d141sm1,d157sm1})//Column
%Out[282]= {{8.04103,0.333918,24.0809}}
%{{8.17914,0.328395,24.9064}}
%{{7.65078,0.345739,22.1288},{765.078,0.429699,1780.5}}
%{{7.80667,0.346247,22.5465},{780.667,0.430894,1811.74}}
%{{4.27389,0.396614,10.7759},{427.389,0.54007,791.358}}

The similarity of the results of PDE solutions (Fig \ref{F:fig3.periodic}, \ref{F:fig3.repellent}) and individual-based simulations (Fig \ref{F:fig3.CPM}) supports this conclusion. This is the more remarkable as the parameters of the PDE model and the cellular Potts model don't correspond. As a result, cellular Potts model simulations are not expected to produce results that agree in quantitative detail to those of the PDE model, not even statistically. In addition, the cellular Potts model is defective as a model of \cel chemotaxis. It would therefore be incorrect to regard the cellular Potts model as a correct model to which the PDE model is a continuum approximation. Both models are wrong, although wrong in different ways---i.e., the PDE model approximates a finite, discontinuous worm population with a continuous function $\rho(t,\vx)$ of space and time, while the cellular Potts model models the worm and its chemotaxis in a biological unrealistic way. To the extent that the results nevertheless agree, we may be reassured that they are not the effects or idiosyncratic characteristics such as being continuum or individual-based.

\section*{Materials and methods}

\subsection*{Numerical solution of partial differential equations (PDEs)}
\label{S:PDEsoln}

We simulated \cel L1 aggregation by solving PDEs (\ref{E:UPDE}, \ref{E:wormrhoPDE}) or (\ref{E:UaPDE}, \ref{E:UrPDE}, \ref{E:wormrhoPDE}) numerically in one or two spatial dimensions. In models with repellent and attractant, there were three PDEs, one for $\rho$ \eqref{E:wormrhoPDE} and one each for $U_a$ \eqref{E:UaPDE}, and $U_r$ \eqref{E:UrPDE} . The domain for one-dimensional simulations was a simple interval $\Omega=[0,w]$. Domains for two-dimensional simulations were rectangular $\Omega=[0,w]\times[0,h]$. Width $w$ and height $h$ varied according to the problem. To avoid distortion of the behavior by boundary effects, all simulations were carried out with periodic boundary conditions. (For an explanation and examples of these boundary effects, see Avery \cite{Avery2020}). 

Continuous fields $\rho$, $U_a$ and $U_r$ were approximated by a grid of points equally spaced in each dimension. The spatial derivatives in the PDEs were replaced with linear combinations of the function values $\rho$, $U_a$, $U_r$, and $V(\rho,U_a,U_r)$ (\ref{E:Vudef}, \ref{E:Vrhodef2}) to approximate the time derivative of each field at each point to fourth order. Simulation of the attractant+repellent model at a resolution of \SI{384}{cm^{-1}} on a \SI{6 x 6}{cm} domain requires $3\times (6\times 384)^2 =$ \num{15925248} degrees of freedom.

We implemented the solution of this system of ODEs (ordinary differential equations) in \petsc (the ``Portable Extensible Toolkit for Scientific computation'') \citeleft\citenum{Balay2017}\citepunct\citenum{petsc-user-ref}\citepunct\citenum{petsc-efficient}\citeright. Among the tools included in \petsc is the \texttt{TS} (time-stepper) package, a library of ODE/DAE (differential algebraic equation) solvers \cite{abhyankar2018}. All solutions shown were produced with the \petsc Rosenbrock-W time stepper \texttt{ra34pw2} \cite{Rang2005}, an implicit third-order method. We used \petsc's basic adaptive step size mechanism. This method uses error estimates from the embedded stepper to adjust step size so as to maintain error below predetermined absolute and relative tolerances. In addition, we imposed a step size limit inspired by the Courant-Friedrichs-Lewy (CFL) condition. At each step we calculated the mean worm velocity $\mathbf{v}=-\grad{V}$ at each point. We limited step size to $\min(\abs{\Delta x/v_x},\abs{\Delta y/v_y})$. (Here $\Delta x$ and $\Delta y$ are the point spacing in the $x$ and $y$ directions, and the minimum is taken over both dimensions and all spatial points.)

Linear equations were solved with the MUMPS parallel direct solver \citeleft\citenum{MUMPS:2}\citepunct\citenum{MUMPS:1}\citeright{} for one-dimensional problems and with \petsc's built-in \texttt{gmres} (generalized minimal residual) Krylov solver for two-dimensional problems. 

\subsection*{Cellular Potts model simulations}

Individual-based model simulations were run in the Morpheus \cite{Starruss2014} modeling environment, using a cellular Potts model \citeleft\citenum{Graner1992}\citepunct\citenum{Glazier1993}\citeright\  to model worm movement. We sought to develop models that corresponded as closely as possible to the PDE model. PDEs describing chemical fields (\ref{E:UPDE}, \ref{E:UaPDE}, \ref{E:UrPDE}) can be reproduced exactly in Morpheus. 

Parameters of the cellular Potts model, unfortunately, do not correspond in any simple way to those that determine worm velocity and dispersal in $\rho$ PDE \eqref{E:wormrhoPDE}. Although valiant efforts have been made to relate the cellular Potts model to continuum models \citeleft\citenum{Alber2007}\citepunct\citenum{Dyson2012}\citeright, these require such drastic simplifying assumptions (e.g. that worms are coordinate axis-aligned rectangles) as to be practically useless. Instead, we calibrated the cellular Potts model by running it with a wide range of parameter values. Perfect calibration is not possible because, for instance, mean velocity varies nonlinearly with the strength of the chemotactic potential gradient. Although this problem can in principle be fixed by taking very small time steps, that would exacerbate an already severe computational feasibility problem. (The simulation shown in Fig \ref{F:fig3.CPM}C,D,E required 9 days. While 9 days is not an unreasonable time to wait for a result if one does everything right the first time, for such sublunary beings as ourselves a \SI{9}{day} run time is a serious inconvenience.)

\subsection*{Full-scale simulations}

Full-scale simulations were carried out on a \SI{6 x 6}{cm} square. The initial condition placed \num{72000} worms on the square, for a final mean density $\bar{\rho} = $ \SI{2000}{cm^{-2}}. (This mean density is much lower than the mean density $\bar{\rho} = $ \SI{9000}{cm^{-2}} used for small-scale simulations with uniform density, because in the full-scale simulations worms are concentrated near the center of the domain, where the density is higher than the mean.) These \num{72000} worms were made up of \num{3600} distributed uniformly on the plate to avoid zero or negative densities, which would result in \eqref{E:wormrhoPDE} becoming undefined, plus \num{68400} worms placed in a \SI{2}{cm} diameter circle at the center of the square, with the density in the square as if a \SI{2}{cm} diameter sphere had been placed in the center and the worms in it fell vertically onto the surface.

\begin{align}
    \label{E:rho0FS}
    \rho(0, x, y) &= b_{\rho} + a_{\rho}\sqrt{\max\qty(0, 1 - \frac{(x-3)^2 + (y-3)^2}{R^2})} \\
    a_{\rho} &= \frac{3(\bar{\rho} - b_{\rho})w^2}{2\pi R^2} \\
    b_{\rho} &= \text{ \SI{100}{cm^{-2}}}\\
    R &= \text{ \SI{1}{cm}} \\
    \bar{\rho} &= \text{ \SI{2000}{cm^{-2}}}\\
    w &= \text{ \SI{6}{cm}}
\end{align}

To this initial condition was added normally distributed pseudorandom noise with standard deviation $0.01\rho(0, x_i, y_j)$ at each grid point $(x_i, y_j)$. In addition, to simulate the continuous production of noise that occurs in real experiments, pseudorandom noise was injected in the course of the simulation. Noise generation was modeled as an independent geometric Brownian motion attached to each grid point. Noise was injected at times $t_n = 10^{n/2}$ {s} for $n$ from \numrange{0}{10}, i.e., at \SIlist{1;3.16;10;31.6;100;316;1000;3162;10000;31622;100000}{s}. To inject noise, $\rho(t_n, x_i, y_j)$ was multiplied by a pseudorandom number $\exp{\mathrm{P}_{n}(x_i, y_j)}$ where $\mathrm{P}_n(x_i, y_j)$ are independent normal pseudorandom variates with variance $10^{-6}\Delta t$, $\Delta t$ being the amount of time passed since the last noise injection. After noise injection, $\rho(t_n, x_i, y_j)$ were normalized so that the total number of worms remained unchanged. The timing of noise injection was a pragmatic compromise. Computational efficiency precludes injecting noise continuously, since time steps had to be small immediately after noise injection---high spatial frequency noise resulted in rapid worm movement. Worms moved rapidly near the beginning of the simulation and more slowly near the end, as they approach a stable equilibrium. We thus chose a schedule in which the frequency of noise injection decreased steadily with time. 

Unfortunately, numerical solution of the cellular Potts model version of the full-scale simulation was computationally infeasible. 

\subsection*{Spectral analysis}

For spectral analysis, the final frame of the video \texttt{N2\_5e5\_washed.mp4} was cropped to a \num{960 x 960} pixel square---this corresponds to a \SI{1.93 x 1.93}{cm} area of the agar surface. The image was standardized so that brightness $b$ varied from  0 to 1, and the discrete Fourier transform calculated with Mathematica \cite{WolframResearch2019} function \texttt{Fourier}. This produces a two-dimensional array of Fourier coefficients $\tilde{b}_\vk$, with wavenumber vector $\vk = (k_x, k_y) \in (2\pi/w)\mathbb{Z}^2$. Power at wavenumber $\vk$ was computed as $p_{\vk} = \abs{\tilde{b}_\vk}^2$. $k_x$ and $k_y$ range from $(-479)\times2\pi/w$ to $480\times2\pi/w$, where $w  = $ \SI{1.93}{cm} is the width of the square. Fig \ref{F:spectralAnalysis}C plots power for coefficients with $k_x/(2\pi)$ ranging from 0 to \SI{20}{cm^{-1}} and $k_y/(2\pi)$ from $-20$ to \SI{20}{cm^{-1}}. We don't show the power for $k_x < 0$ because the power spectrum is even in $\vk$, i.e. $p_{-\vk} = p_\vk$. $k/(2\pi)$ is the inverse of the wavelength of the corresponding sinusoid, so, for instance, $k_x/(2\pi) = $ \SI{10}{cm^{-1}} corresponds to a sinusoid whose wavelength in the $x$ direction is \SI{1}{mm}. 

For comparison to experiment a \SI{200000}{s} image of the full-scale simulation was cropped and scaled and standardized to a $[0,1]$ range. The cropping square was chosen so that the center point of the simulation was located at the same place as the center of the central mass in the cropped experimental image. The scale was chosen to make the major peaks in the radial power spectrum (Fig \ref{F:spectralAnalysis}F) near $2\pi\times$\SI{9.5}{cm^{-1}} correspond precisely. Finally a black circle of the same dimensions as the central mass in the video image was added.

The radially summed power spectrum was approximated by an array $s$ of 1024 numbers representing power at wavenumbers from 0 to $2\pi\times20$ \si{cm^{-1}}. Each wavenumber in the two-dimensional power spectrum was mapped linearly to a point in $s$. Since most wavenumbers mapped to noninteger locations, they were represented by a weighted sum of the two closest elements of $s$. For instance, consider $\vk = 2\pi\times (1, 2)/1.93$. This corresponds in the radial spectrum to wavenumber $k = 2\pi\times\sqrt{5}/1.93 = 2\pi\times 1.15$ \si{cm^{-1}}. This $k$ maps to radial array $s$ location $j = 1 + 1023(1.15/20) = 60.26$. If $p_{\vk}$ is the power at $\vk$ in the two-dimensional spectrum, we add $(61-j)p_{\vk}$ to $s[60]$ and $(j-60)p_{\vk}$ to $s[61]$. 

The radially summed spectrum thus computed is noisy and quasi-periodic with period $1023/(20\times 1.93)\approx 26.5$, reflecting the discrete two-dimensional power spectrum. To produce Fig \ref{F:spectralAnalysis}F, we smoothed $s$ using Mathematica function \texttt{GaussianFilter} with smoothing radius $26.5/\sqrt{2}\approx 18.7$ which we determined by trial and error to be the smallest smoothing radius that effectively eliminated the periodic structure. 

\subsection*{Effect of ethanol and acetate on the transcriptome of starved L1s}

To obtain L1s for transcriptomic profiling, we grew N2 worms in liquid culture. We inoculated \SI{250}{mL} S-complete in a \SI{2}{L} flask with \num{7e5} synchronized L1 larvae obtained from a small-scale liquid culture and added \SI{10}{mL} 50\% \textit{E. coli} K-12 stock suspension. Worms were grown at 22\textdegree C, 220 rpm (for details see Artyukhin et al \cite{Artyukhin2015}). We monitored the worm culture during the next \SI{2.5}{days} and added \textit{E. coli} food as it became depleted. Bleaching of gravid adult worms (\SI{8}{mL} water + \SI{2}{mL} bleach + \SI{0.3}{mL} \SI{10}{M} NaOH for \SI{6}{min}) after \SI{68}{h} of growth yielded ca. \num{1e7} eggs. After 3 washes with M9 buffer, eggs were resuspended in \SI{20}{mL} M9 and allowed to hatch. After 2 days of starvation (\SIrange{51}{53}{h} after bleaching, 20\textdegree C), we collected L1 larvae, washed them 6 times with M9, \SI{10}{mL} each time, and resuspended in \SI{10}{mL} M9 after the final wash. \SI{300}{\micro\liter} of the resulting L1 suspension (corresponding to ca. \SI{20}{\micro\liter} L1 pellet) were added to each of the following solutions in 15-ml plastic tubes: (1) \SI{3}{mL} M9; (2) \SI{3}{mL} M9 + \SI{3}{\micro\liter} ethanol; (3) \SI{3}{mL} M9 + \SI{51}{\micro\liter} \SI{1}{M} potassium acetate. Tubes were put on a rocker at room temperature (ca. 21\textdegree C). After \SI{1.5}{h}, the tubes were cooled down on ice for \SI{1}{\min}, centrifuged, and ca. \SI{2}{mL} of supernatant was removed by aspiration. We resuspended L1s in the remaining ca. \SI{1}{mL} of liquid, transferred the suspension to \SI{1.5}{mL} tubes, centrifuged, removed supernatant, added \SI{300}{\micro\liter} Trizol to each sample, and froze in liquid nitrogen. The above procedure was repeated two more times with worms grown on different days to obtain biological triplicates for each condition (control, ethanol, potassium acetate). All samples were stored at -80\textdegree C before analysis.

\subsection*{Microarray methods}

Microarray methods were as described by Hyun et al \cite{Hyun2016}.

\subsection*{\srhtwo knockout}

Deletion mutants of \srhtwo were generated by CRISPR and obtained from Knudra. Three deletion strains were generated (COP-1274, 1275, 1276), all of the same genotype: \textit{unc--119(ed3) III; srh--2(knu317::unc--119(+)) V}. 

\section*{Acknowledgements}

We thank Leah Edelstein-Keshet for introducing us to the Morpheus modeling environment.

\nolinenumbers

% Either type in your references using
% \begin{thebibliography}{}
% \bibitem{}
% Text
% \end{thebibliography}
%
% or
%
% Compile your BiBTeX database using our plos2015.bst
% style file and paste the contents of your .bbl file
% here. See http://journals.plos.org/plosone/s/latex for 
% step-by-step instructions.
% 

%\bibliography{references,citation}
%\bibliography{references}

\section*{References}

\begin{thebibliography}{10}

\bibitem{Barr2006}
Barr MM, Garcia LR. {Male mating behavior.}; 2006.

\bibitem{Artyukhin2015}
Artyukhin AB, Yim JJ, Cheong MC, Avery L.
\newblock {Starvation-induced collective behavior in C. elegans.}
\newblock Scientific reports. 2015;5:10647.
\newblock doi:{10.1038/srep10647}.

\bibitem{DeBono1998}
de~Bono M, Bargmann CI.
\newblock {Natural variation in a neuropeptide Y receptor homolog modifies
  social behavior and food response in C. elegans}.
\newblock Cell. 1998;94(5):679--689.

\bibitem{DeBono2002}
de~Bono M, Tobin DM, Davis MW, Avery L, Bargmann CI.
\newblock {Social feeding in Caenorhabditis elegans is induced by neurons that
  detect aversive stimuli}.
\newblock Nature. 2002;419(6910):899--903.

\bibitem{Rogers2006}
Rogers C, Persson A, Cheung B, de~Bono M.
\newblock {Behavioral motifs and neural pathways coordinating O2 responses and
  aggregation in C. elegans}.
\newblock Curr Biol. 2006;16(7):649--659.
\newblock doi:{S0960-9822(06)01316-9 [pii] 10.1016/j.cub.2006.03.023}.

\bibitem{Ding2019}
Ding SS, Schumacher LJ, Javer AE, Endres RG, Brown AE.
\newblock {Shared behavioral mechanisms underlie C. elegans aggregation and
  swarming}.
\newblock eLife. 2019;8.
\newblock doi:{10.7554/eLife.43318}.

\bibitem{Demir2020}
Demir E, Yaman YI, Basaran M, Kocabas A.
\newblock {Dynamics of pattern formation and emergence of swarming in c.
  Elegans}.
\newblock eLife. 2020;9.
\newblock doi:{10.7554/eLife.52781}.

\bibitem{VonReuss2012a}
Von~Reuss SH, Bose N, Srinivasan J, Yim JJ, Judkins JC, Sternberg PW, et~al.
\newblock {Comparative metabolomics reveals biogenesis of ascarosides, a
  modular library of small-molecule signals in C. elegans}.
\newblock Journal of the American Chemical Society. 2012;134(3):1817--1824.
\newblock doi:{10.1021/ja210202y}.

\bibitem{Butcher2009}
Butcher RA, Ragains JR, Li W, Ruvkun G, Clardy J, Mak HY.
\newblock {Biosynthesis of the Caenorhabditis elegans dauer pheromone}.
\newblock Proc Natl Acad Sci U S A. 2009;106(6):1875--1879.
\newblock doi:{0810338106 [pii] 10.1073/pnas.0810338106}.

\bibitem{Golden1985}
Golden JW, Riddle DL.
\newblock {A gene affecting production of the Caenorhabditis elegans
  dauer-inducing pheromone}.
\newblock Mol Gen Genet. 1985;198(3):534--536.

\bibitem{Joo2009}
Joo HJ, Yim YH, Jeong PY, Jin YX, Lee JE, Kim H, et~al.
\newblock {Caenorhabditis elegans utilizes dauer pheromone biosynthesis to
  dispose of toxic peroxisomal fatty acids for cellular homoeostasis.}
\newblock The Biochemical journal. 2009;422(1):61--71.
\newblock doi:{10.1042/BJ20090513}.

\bibitem{Artyukhin2015a}
Artyukhin AB, Yim JJ, Cheong~Cheong M, Avery L.
\newblock {Starvation-induced collective behavior in C. elegans}.
\newblock Scientific Reports. 2015;5.
\newblock doi:{10.1038/srep10647}.

\bibitem{Sugi2019}
Sugi T, Ito H, Nishimura M, Nagai KH.
\newblock {C. elegans collectively forms dynamical networks}.
\newblock Nature Communications. 2019;10(1):683.
\newblock doi:{10.1038/s41467-019-08537-y}.

\bibitem{Bargmann2006}
Bargmann CI.
\newblock {Chemosensation in C. elegans}.
\newblock WormBook. 2006; p. 1--29.
\newblock doi:{10.1895/wormbook.1.123.1}.

\bibitem{Bargmann1997}
Bargmann CI, Mori I.
\newblock {Chemotaxis and Thermotaxis}.
\newblock In: Riddle DL, Blumenthal T, Meyer BJ, Preiss JR, editors. C. elegans
  II. New York: Cold Spring Harbor Press; 1997. p. 717--737.

\bibitem{Mori1999}
Mori I.
\newblock {Genetics of chemotaxis and thermotaxis in the nematode
  Caenorhabditis elegans}.
\newblock Annu Rev Genet. 1999;33:399--422.

\bibitem{Dunn2004}
Dunn NA, Lockery SR, Pierce-Shimomura JT, Conery JS.
\newblock {A Neural Network Model of Chemotaxis Predicts Functions of Synaptic
  Connections in the Nematode Caenorhabditis elegans}.
\newblock Journal of Computational Neuroscience. 2004;17(2):137--147.
\newblock doi:{10.1023/B:JCNS.0000037679.42570.d5}.

\bibitem{Roberts2016}
Roberts WM, Augustine SB, Lawton KJ, Lindsay TH, Thiele TR, Izquierdo EJ,
  et~al.
\newblock {A stochastic neuronal model predicts random search behaviors at
  multiple spatial scales in C. elegans}.
\newblock eLife. 2016;5.
\newblock doi:{10.7554/eLife.12572}.

\bibitem{Pierce-Shimomura1999}
Pierce-Shimomura JT, Morse TM, Lockery SR.
\newblock {The fundamental role of pirouettes in Caenorhabditis elegans
  chemotaxis.}
\newblock The Journal of neuroscience : the official journal of the Society for
  Neuroscience. 1999;19(21):9557--69.
\newblock doi:{10.1523/JNEUROSCI.19-21-09557.1999}.

\bibitem{Ferree2020}
Ferr{\'{e}}e TC, Lockery SR.
\newblock {Computational rules for chemotaxis in the nematode C. elegans.}
\newblock Journal of computational neuroscience. 2020;6(3):263--77.
\newblock doi:{10.1023/a:1008857906763}.

\bibitem{Keller1970}
Keller EF, Segel LA.
\newblock {Initiation of slime mold aggregation viewed as an instability}.
\newblock Journal of Theoretical Biology. 1970;26(3):399--415.
\newblock doi:{10.1016/0022-5193(70)90092-5}.

\bibitem{Edelstein-Keshet2005}
Edelstein-Keshet L.
\newblock {Mathematical models in biology}.
\newblock Society for Industrial and Applied Mathematics; 2005.

\bibitem{Horstmann2003}
Horstmann D.
\newblock {From 1970 until present: the Keller-Segel model in chemotaxis and
  its consequences. I.}
\newblock Jahresber Deutsch Math-Verein. 2003;105(3):103–165.

\bibitem{Horstmann2004}
Horstmann D.
\newblock {From 1970 until present: The Keller-Segel model in chemotaxis and
  its consequences II.}
\newblock Jahresber Deutsch Math-Verein. 2004;106(2):51--69.

\bibitem{Hillen2009}
Hillen T, Painter KJ.
\newblock {A user’s guide to PDE models for chemotaxis}.
\newblock Journal of Mathematical Biology. 2009;58(1-2):183--217.
\newblock doi:{10.1007/s00285-008-0201-3}.

\bibitem{Avery2020}
{Avery, Leon}.
\newblock Mathematical Modeling of C elegans L1 aggregation [Ph.D. thesis].
\newblock University of Waterloo; 2020.
\newblock Available from: \url{http://hdl.handle.net/10012/15480}.

\bibitem{Gescheider1997}
Gescheider GA.
\newblock {Psychophysics: The Fundamentals: Gescheider, George A.:
  9781138984158: Cognitive Psychology: Amazon Canada}.
\newblock Third edit ed. Psychology Press; 1997.

\bibitem{Avery2003}
Avery L, Shtonda BB.
\newblock {Food transport in the C. elegans pharynx}.
\newblock J Exp Biol. 2003;206(Pt 14):2441--2457.

\bibitem{Starruss2014}
Starruss J, de~Back W, Brusch L, Deutsch A.
\newblock {Morpheus: a user-friendly modeling environment for multiscale and
  multicellular systems biology}.
\newblock Bioinformatics. 2014;30(9):1331--1332.
\newblock doi:{10.1093/bioinformatics/btt772}.

\bibitem{Luca2003}
Luca M, Chavez-Ross A, Edelstein-Keshet L, Mogilner A.
\newblock {Chemotactic signaling, microglia, and Alzheimer's disease senile
  plaques: Is there a connection?}
\newblock Bulletin of Mathematical Biology. 2003;65(4):693--730.
\newblock doi:{10.1016/S0092-8240(03)00030-2}.

\bibitem{Mogilner2003}
Mogilner A, Edelstein-Keshet L, Bent L, Spiros A.
\newblock {Mutual interactions, potentials, and individual distance in a social
  aggregation}.
\newblock Journal of Mathematical Biology. 2003;47(4):353--389.
\newblock doi:{10.1007/s00285-003-0209-7}.

\bibitem{Balay2017}
Balay S, Abhyankar S, Adams MF, Brown J, Gropp P, Buschelman K, et~al.. {PETSc
  Web Page}; 2017.

\bibitem{petsc-user-ref}
Balay S, Abhyankar S, Adams MF, Brown J, Brune P, Buschelman K, et~al.
\newblock {{\{}PETS{\}}c Users Manual}.
\newblock Argonne National Laboratory; 2019. ANL-95/11 - Revision 3.11.
\newblock Available from: \url{https://www.mcs.anl.gov/petsc}.

\bibitem{petsc-efficient}
Balay S, Gropp WD, McInnes LC, Smith BF.
\newblock {Efficient Management of Parallelism in Object Oriented Numerical
  Software Libraries}.
\newblock In: Arge E, Bruaset AM, Langtangen HP, editors. Modern Software Tools
  in Scientific Computing. Birkh{\"{a}}user Press; 1997. p. 163--202.

\bibitem{abhyankar2018}
Abhyankar S, Brown J, Constantinescu EM, Ghosh D, Smith BF, Zhang H.
\newblock {PETSc/TS: A Modern Scalable ODE/DAE Solver Library}.
\newblock arXiv preprint arXiv:180601437. 2018;.

\bibitem{Rang2005}
Rang J, Angermann L.
\newblock {New Rosenbrock W-methods of order 3 for partial differential
  algebraic equations of index 1}.
\newblock BIT Numerical Mathematics. 2005;45(4):761--787.

\bibitem{MUMPS:2}
Amestoy PR, Guermouche A, L'Excellent JY, Pralet S.
\newblock {Hybrid scheduling for the parallel solution of linear systems}.
\newblock Parallel Computing. 2006;32(2):136--156.

\bibitem{MUMPS:1}
Amestoy PR, Duff IS, Koster J, L'Excellent JY.
\newblock {A Fully Asynchronous Multifrontal Solver Using Distributed Dynamic
  Scheduling}.
\newblock SIAM Journal on Matrix Analysis and Applications. 2001;23(1):15--41.

\bibitem{Graner1992}
Graner F, Glazier JA.
\newblock {Simulation of biological cell sorting using a two-dimensional
  extended Potts model}.
\newblock Physical Review Letters. 1992;69(13):2013--2016.
\newblock doi:{10.1103/PhysRevLett.69.2013}.

\bibitem{Glazier1993}
Glazier JA, Graner F.
\newblock {Simulation of the differential adhesion driven rearrangement of
  biological cells}.
\newblock Physical Review E. 1993;47(3):2128--2154.
\newblock doi:{10.1103/PhysRevE.47.2128}.

\bibitem{Alber2007}
Alber M, Chen N, Lushnikov PM, Newman SA.
\newblock {Continuous macroscopic limit of a discrete stochastic model for
  interaction of living cells}.
\newblock Physical Review Letters. 2007;99(16):168102.
\newblock doi:{10.1103/PhysRevLett.99.168102}.

\bibitem{Dyson2012}
Dyson L, Maini PK, Baker RE.
\newblock {Macroscopic limits of individual-based models for motile cell
  populations with volume exclusion}.
\newblock Physical Review E - Statistical, Nonlinear, and Soft Matter Physics.
  2012;86(3).
\newblock doi:{10.1103/PhysRevE.86.031903}.

\bibitem{WolframResearch2019}
Wolfram~Research I. {Mathematica}; 2019.

\bibitem{Hyun2016}
Hyun M, Davis K, Lee I, Kim J, Dumur C, You YJ.
\newblock {Fat Metabolism Regulates Satiety Behavior in C. elegans}.
\newblock Scientific Reports. 2016;6.
\newblock doi:{10.1038/srep24841}.

\end{thebibliography}
%\begin{thebibliography}{10}

%\begin{thebibliography}{10}
%
%\bibitem{bib1}
%Conant GC, Wolfe KH.
%\newblock {{T}urning a hobby into a job: how duplicated genes find new
%  functions}.
%\newblock Nat Rev Genet. 2008 Dec;9(12):938--950.
%
%\bibitem{bib2}
%Ohno S.
%\newblock Evolution by gene duplication.
%\newblock London: George Alien \& Unwin Ltd. Berlin, Heidelberg and New York:
%  Springer-Verlag.; 1970.
%
%\bibitem{bib3}
%Magwire MM, Bayer F, Webster CL, Cao C, Jiggins FM.
%\newblock {{S}uccessive increases in the resistance of {D}rosophila to viral
%  infection through a transposon insertion followed by a {D}uplication}.
%\newblock PLoS Genet. 2011 Oct;7(10):e1002337.
%
%\end{thebibliography}

\section*{Supporting information}

\beginsupplement

\subsection*{Linear stability analysis of the attractant-only model}

The attractant--only model (\ref{E:UPDE}, \ref{E:wormrhoPDE}) shares with the original Keller-Segel system \cite{Keller1970} the property of density-dependent instability. There is a uniform equilibrium,

\begin{align}
    \label{E:rhoeq}
    \rho_\text{eq}(t, \vx) &= \bar{\rho} \\
    \label{E:Ueq}
    U_\text{eq}(t, \vx) &= \bar{U} \coloneqq \frac{s_a}{\gamma_a}\bar{\rho}
\end{align}

\noindent
Substituting these functions into (\ref{E:UPDE}, \ref{E:wormrhoPDE}) shows that $\dv*{\rho_\text{eq}}{t} = \dv*{U_\text{eq}}{t} = 0$. Consider a population near this equilibrium, and let $\rho(t,\vx) = \bar{\rho} + \delta\rho(t,\vx)$, $U(t,\vx) = \bar{U} + \delta U(t,\vx)$. Because $(\bar{\rho}, \bar{U})$ is an equilibrium, $\rho_t(t,\vx) = \delta\rho_t(t,\vx)$ and $U_t(t,\vx) = \delta U_t(t,\vx)$. Substituting into (\ref{E:UPDE}, \ref{E:wormrhoPDE}), 

\begin{align}
    \delta\rho_t(t,\vx) &= \div\left(\delta\rho V_U'(U)\grad \bar{U} +\bar{\rho}V_U'(U)\grad\delta U\right) \\
    &\quad+ \div\qty(\delta\rho V_\rho'(\rho)\grad\bar{\rho} + \bar{\rho}V_\rho'(\rho)\grad\delta\rho) \notag \\
    &\quad+ \div\left(\delta\rho V_U'(U) \grad\delta U\right) \notag \\ \
    &\quad+ \sigma\laplacian\bar{\rho} + \sigma\laplacian\delta\rho \notag\\
    \delta U_t(t,\vx) &= 
    -\gamma\delta U + D\laplacian\delta U+s\delta\rho
\end{align}

\noindent
$\grad \bar{U} = \grad\bar{\rho} = \laplacian\bar{\rho} = 0$. Writing $V_U'(U)=V_U'(\bar{U})+\order{\delta U}$, $V_\rho'(\rho)=V_\rho'(\bar{\rho})+\order{\delta \rho}$ and ignoring second order terms we have, to first order, the linear vector PDE, 

\begin{equation}
    \label{E:linPDE}
    \frac{\partial}{\partial t}\left(
    \begin{matrix}
        \delta\rho \\
        \delta U
    \end{matrix}
    \right) = \left(
    \begin{matrix}
        (\sigma+V_\rho'(\bar{\rho}))\nabla^2 & V_U'(\bar{U})\bar{\rho}\nabla^2 \\
        s              & -\gamma + D\nabla^2
    \end{matrix}
    \right)
    \left(
    \begin{matrix}
        \delta\rho \\
        \delta U
    \end{matrix}
    \right)
\end{equation}

\noindent
This is easily solved by separation of variables. It will be convenient to define $\sigma' = \sigma+V_\rho'(\bar{\rho})$. Since $V_\rho$ is, by design, an increasing function, $V_\rho'>0$ and $\sigma' > \sigma > 0$. For the parameters in Table \ref{T:ch3.params}, $V_\rho'(\bar{\rho})\approx 0$ and $\sigma' \approx \sigma$. Now, eigenfunctions of \eqref{E:linPDE} are of the form 

\begin{equation}
    \label{E:linEfuncs}
    \left(
    \begin{matrix}
        \delta\rho(t,\vx) \\
        \delta U(t,\vx)
    \end{matrix}
    \right) = \left(
    \begin{matrix}
        \rho_\vk e^{\lambda_\vk t}e^{i\vk\cdot\vx} \\
        U_\vk e^{\lambda_\vk t}e^{i\vk\cdot\vx}
    \end{matrix}
    \right)
\end{equation}

\noindent 
$\vk$ is a wavenumber vector, i.e., a vector of frequency in each spatial direction. Substituting into \eqref{E:linPDE} produces the $2\times2$ matrix eigenvalue problem,

\begin{equation}
\label{E:linmat}
    \lambda_\vk\left(
    \begin{matrix}
        \rho_\vk \\
        U_\vk
    \end{matrix}
    \right) = \left(
    \begin{matrix}
        -\sigma'k^2 & -V_U'(\bar{U})\bar{\rho} k^2 \\
        s           & -\gamma - D k^2
    \end{matrix}
    \right)\left(
    \begin{matrix}
        \rho_\vk \\
        U_\vk
    \end{matrix}
    \right)
\end{equation}

\noindent 
where $k^2 \coloneqq \norm{\vk}^2$. Solutions of the linearized system \eqref{E:linmat} are linear combinations of functions \eqref{E:linEfuncs} where $(\rho_\vk,U_\vk)^\intercal$ is an eigenvector of the matrix in \eqref{E:linmat}. If, for every $\vk$, $\Re(\lambda_\vk) < 0$ then any small fluctuation away from the uniform equilibrium will die away, and the uniform equilibrium will be stable. If, however, there exists a $\vk$ such that the matrix has an eigenvalue with positive real part, then the uniform equilibrium is unstable. The sum of the two eigenvalues, the trace of the matrix, is negative, $-\sigma'k^2-\gamma - D k^2 < 0$, so the only possible way to have an eigenvalue with positive real part is if both eigenvalues are real, one positive and one negative. Thus, first-order instability is expected if and only if the determinant is negative,

\begin{align}
\label{E:determinant}
    \left\lvert\begin{matrix}
        -\sigma'k^2 & -V_U'(\bar{U})\bar{\rho} k^2 \\
        s           & -\gamma - D k^2
    \end{matrix}\right\rvert < 0 \\
    \label{E:instab1}
    D\sigma'k^4 + \left(\gamma \sigma' + V_U'(\bar{U}) s\bar{\rho}\right)k^2 < 0
\end{align}

Inequality \eqref{E:instab1} can hold only if $\gamma\sigma' + V_U'(\bar{U}) s\bar{\rho} < 0$. Since, as mentioned above, $V_U$ is a decreasing function of attractant concentration, $V_U'(\bar{U}) < 0$. If the condition $\gamma\sigma' + V_U'(\bar{U}) s\bar{\rho} < 0$ holds, then for some small enough $k^2$, the negative $k^2$ term will dominate the positive $k^4$ term and the determinant will be negative. Thus, first-order instability occurs if and only if $\bar{\rho} > -\frac{\gamma\sigma'}{s V_U'(\bar{U})}$. Since $\bar{U} \coloneqq s\bar{\rho}/\gamma$ depends on $\bar{\rho}$, it is not a foregone conclusion that instability is possible. Neglecting $V_\rho'$ and with $V_U$ as in \eqref{E:Vudef}, the instability condition reduces to

\begin{equation}
    \label{E:instcond}
    \bar{\rho} > \frac{\alpha\gamma\sigma}{s(\beta-\sigma)} > 0
\end{equation}

\noindent
Instability is possible if $\beta>\sigma$. By design, the instability condition is $\bar{\rho} > $\SI{1500}{cm^{-d}} with the parameter values in Table \ref{T:ch3.params}.

\subsection*{Linear stability analysis of the attractant+repellent model}

For simplicity, we assume $V_\rho = 0$ in the following analysis. With the parameter values in Table \ref{T:ch3.params}, this is very close to true in the vicinity of the instability threshold. 
Linearization of the PDE system (\ref{E:wormrhoPDE}, \ref{E:UaPDE}, \ref{E:UrPDE}) around a uniform equilibrium at $\rho_\text{eq}(t, \vx) = \bar{\rho}$, $U_{i,\text{eq}}(t, \vx) = \bar{U_i} = s_i\bar{\rho}/\gamma_i$ (for $i \in a,r$) produces the following linear PDE system

\begin{equation}
    \label{E:arlinPDE}
    \dv{t}\begin{pmatrix}
        \delta\rho \\
        \delta U_a \\
        \delta U_r
    \end{pmatrix} =
    \begin{pmatrix}
        \sigma \nabla^2 & \bar{\rho} V'_{U_a}(\bar{U_a}) \nabla^2 & \bar{\rho} V'_{U_r}(\bar{U_r}) \nabla^2 \\
        s_a & -\gamma_a + D_a\nabla^2 & 0 \\
        s_r & 0 & -\gamma_r + D_r\nabla^2 \\ 
    \end{pmatrix}
    \begin{pmatrix}
        \delta\rho \\
        \delta U_a \\
        \delta U_r
    \end{pmatrix}
\end{equation}

\noindent 
The ansatz

\begin{equation}
\label{E:linansatz}
    \begin{pmatrix}
        \delta\rho \\
        \delta U_a \\
        \delta U_r
    \end{pmatrix}
    = \begin{pmatrix}
        \rho_\vk \\
        U_{a\vk} \\
        U_{r\vk}
    \end{pmatrix} e^{\lambda_\vk t} e^{i\vk\cdot\vx}
\end{equation}

\noindent 
yields the eigenvalue problem

\begin{align}
    \label{E:armatrix1}
    \lambda_\vk\begin{pmatrix}
        \rho_\vk \\
        U_{a\vk} \\
        U_{r\vk}
    \end{pmatrix} &=\begin{pmatrix}
        -\sigma k^2 & -\bar{\rho} V'_{U_a}(\bar{U_a}) k^2 & -\bar{\rho} V'_{U_r}(\bar{U_r}) k^2 \\
        s_a & -\gamma_a - D_a k^2 & 0 \\
        s_r & 0 & -\gamma_r - D_r k^2 \\
    \end{pmatrix}
    \begin{pmatrix}
        \rho_\vk \\
        U_{a\vk} \\
        U_{r\vk}
    \end{pmatrix} \\
    \label{E:blkarmat1}
    &=\begin{pmatrix}
        -\sigma k^2 & -\bar{\rho} k^2 (\vb{V'}(\vb{\bar{U}}))^\intercal \\
        \bar{\rho}\vb{s} & -\boldsymbol{\Gamma} - \vb{D} k^2 \\
    \end{pmatrix}
    \begin{pmatrix}
        \rho_\vk \\
        \vb{U}_\vk
    \end{pmatrix} \\
    \label{E:narmat1}
    &= -\vb{N}(\bar{\rho},k^2)\begin{pmatrix}
        \rho_\vk \\
        \vb{U}_\vk
    \end{pmatrix}
\end{align}

\noindent 
In \eqref{E:blkarmat1}, the matrix is in a block form that can be extended easily to any number of signals, with 

\begin{align}
    \label{E:armatblks}
    \vb{s} &= \begin{pmatrix}
        s_a \\
        s_r
    \end{pmatrix} \\
    \vb{U}_\vk &= \begin{pmatrix}
        U_{a\vk} \\
        U_{r\vk}
    \end{pmatrix} \\
    \vb{V'}(\vb{\bar{U}}) &= \begin{pmatrix}
        V'_{U_a}(\bar{U_a}) \\
        V'_{U_r}(\bar{U_r})
    \end{pmatrix} \\
    \boldsymbol{\Gamma} &= \begin{pmatrix}
        \gamma_a & 0 \\
        0 & \gamma_r
    \end{pmatrix} \\
    \vb{D} &= \begin{pmatrix}
        D_a & 0 \\
        0 & D_r
    \end{pmatrix}
\end{align}

\noindent 
In \eqref{E:narmat1}, $\vb{N}(\bar{\rho},k^2)$ is defined as the negative of the matrix in \eqref{E:blkarmat1}. (We define $\vb{N}$ as the negative to avoid an inconvenient factor of $(-1)^{1+n_\text{signals}}$ in the determinant we are about to calculate.) The uniform equilibrium is unstable at mean density $\bar{\rho}$ if for some $k$, $\vb{N}(\bar{\rho},k^2)$ has an eigenvalue with negative real part. If $\abs{\vb{N}(\bar{\rho},k^2)} < 0$, the equilibrium is certainly unstable. This leads to the following criterion for instability

\begin{align}
    \label{E:arinstab}
    -\sigma &> \bar{\rho} (\vb{V'}(\vb{\bar{U}}))^\intercal(\boldsymbol{\Gamma}+\vb{D}k^2)^{-1}\vb{s} \\
    \label{E:arinstab2}
    &= \bar{\rho}\sum_{i\in\{a,r\}}\frac{\vb{V'}_{U_i}(\bar{U_i})s_i}{\gamma_i + D_i k^2} \\
    \label{E:arinstab3}
    &= \bar{\rho}\qty(
        \frac{\vb{V'}_{U_a}(\bar{U_a})s_a}{\gamma_a + D_a k^2}
        +\frac{\vb{V'}_{U_r}(\bar{U_r})s_r}{\gamma_r + D_r k^2}
    )
\end{align}

It is possible to choose parameter values so that this criterion predicts instability with a nontrivial minimum wavenumber (and therefore finite maximum scale). How does this work? Remember that $V'_{U_r}>0$ because it is a repellent and $V'_{U_a}<0$ because it is an attractant. Thus the two terms in \eqref{E:arinstab3} are opposite in sign. Also, $\gamma_r < \gamma_a$ and $D_r > D_a$, because the repellent is a longer-range signal than the attractant. Thus, at low $k$, if the relative magnitudes of $V'_{U_r}s_r$ and $V'_{U_a}s_a$ are appropriately adjusted (by evolution or the modeler), the sum in \eqref{E:arinstab3} is positive and the uniform equilibrium is stable to perturbations of small wavenumber = large scale. As $k$ rises the $D_r k^2$ factor in the denominator of the repellent term makes the positive term small compared to the negative attractant term. The sum in parentheses can become negative, and if $\bar{\rho}$ is large enough, the right-hand-side drops below $-\sigma$, and instability to perturbations of intermediate wavenumber = medium scale results. For large $k$ the right-hand-side approaches zero because of the $D k^2$ factors in both denominators. The uniform equilibrium is thus stable to perturbations of large wavenumber = small scale. It is therefore possible for an attractant+repellent Keller-Segel model to have a finite natural scale. 

Based on calculations of this sort we chose $\beta_r = -\beta_a = -2\sigma = $ \SI{-5.56e-4}{cm^2 s^{-1}}. Attractant parameters remained as in Table \ref{T:ch3.params}. The addition of a repellent increases the threshold for instability, but the uniform equilibrium is still predicted to be unstable at $\bar{\rho} = $ \SI{9000}{cm^-d}. 

\subsection*{Convergence tests}
\label{S:convtest}

To test whether the numerical solution of the worm system PDEs (\ref{E:wormrhoPDE}, \ref{E:UaPDE}, \ref{E:UrPDE}) approximates the correct solution, we compared numerical solutions of the attractant+repellent system to an analytical solution of the linearized PDE system \eqref{E:arlinPDE}. In one dimension, for $x\in\Omega=[0,1]$,

\begin{align}
    \label{E:CTsolndrho1D}
    \delta\rho(t, x) &\coloneqq a_\rho e^{\lambda t} \sin(\phi + 2\pi k_0 x) \\
    \label{E:CTsolnrho1D}
    \rho_\text{lin}(t, x) &= \bar{\rho} + \delta\rho(t, x) \\
    \label{E:CTsolnUa1D}
    U_{a,\text{lin}}(t, x) &= \frac{s_a}{\gamma_a}\bar{\rho} + U_{a,k_0} \delta\rho(t, x) \\
    \label{E:CTsolnUr1D}
    U_{r,\text{lin}}(t,x) &= \frac{s_r}{\gamma_r}\bar{\rho} + U_{r,k_0} \delta\rho(t, x)
\end{align}

In two dimensions, for $(x, y)\in\Omega = [0, \sqrt{3}/2]\times[0, 1/2]$, 

\begin{align}
    \label{E:CTsolndrho2D}
    \delta\rho(t, x, y) &\coloneqq a_\rho e^{\lambda t}\bigg(\frac{1}{3}\big(
            \cos(2 \pi k_0 y)+ \\
            &\qquad\qquad\qquad\cos(2 \pi (k_{x,0} x-k_{y,0} y))+ \notag \\
            &\qquad\qquad\qquad\cos(2 \pi (k_{x,0} x+k_{y,0} y))
        \big) \bigg) \notag \\
    \label{E:CTsolnrho2D}
    \rho_\text{lin}(t, x, y) &= \bar{\rho} + \delta\rho(t, x, y) \\
    \label{E:CTsolnUa2D}
    U_{a,\text{lin}}(t, x, y) &= \frac{s_a}{\gamma_a}\bar{\rho} + U_{a,k_0} \delta\rho(t, x, y) \\
    \label{E:CTsolnUr2D}
    U_{r,\text{lin}}(t, x, y) &= \frac{s_r}{\gamma_r}\bar{\rho} + U_{r,k_0} \delta\rho(t, x, y)
\end{align}

\noindent
Here $\phi, a_\rho\in\mathbb{R}$ and $k_0\in 2\mathbb{Z}$ are parameters that can be freely chosen. We chose $a_\rho=1$ and $\phi=\pi/2$. We chose $k_0=4$ to produce a substantial positive growth rate. For the two-dimensional case, we chose $k_{x,0} = k_0\sqrt{3}/2$ and $k_{y,0} = k_0/2$ to produce hexagonal symmetry. $\lambda$ is the positive eigenvalue of matrix $-\vb{N}(\bar{\rho}, k^2_0)$ \eqref{E:narmat1}, with corresponding eigenvector $\qty(1, U_{ak_0}, U_{rk_0})^\intercal$ (as in \eqref{E:linansatz}, but normalized so that $\rho_{k_0} = 1$). Numerical values $\lambda \approx $ \SI{0.000955}{s^{-1}}, $U_{ak_0} \approx$ \num{0.863}, $U_{rk_0} \approx $ \num{0.121} were estimated to 15-digit precision by numerical diagonalization of the computed matrix $-\vb{N}(\bar{\rho}, k^2_0)$. 

Functions $\qty(\rho_\text{lin}(t,x), U_{a,\text{lin}}(t,x), U_{r,\text{lin}}(t,x))^\intercal$ are of course not an exact solution of the full nonlinear PDEs (\ref{E:wormrhoPDE}, \ref{E:UaPDE}, \ref{E:UrPDE}). To produce a closely related system with this exact solution for convergence testing, we modified the $\rho$ PDE \eqref{E:wormrhoPDE} by addition of a source term $S(t,\vx)$.

\begin{align}
    \label{E:convrhoPDE}
    \dot{\rho} &= \nabla\cdot\qty(
        \rho\nabla\qty(
            V_{U_a}(U_a) + V_{U_r}(U_r) + V_\rho(\rho) + \sigma\log\rho
        )) + S(t, \vx) \\
    \label{E:sourcecomp}
    S(t, \vx) &= \pdv{\rho_\text{lin}(t,\vx)}{t}
    - \div{\big(
        \rho_\text{lin}(t,\vx) \\
            &\qquad \grad\big(V_{U_a}(U_{a,\text{lin}}(t,\vx)) + V_{U_r}(U_{r,\text{lin}}(t,\vx)) \notag \\
            &\qquad + V_\rho(\rho_\text{lin}(t,\vx)) \notag \\
            &\qquad + \sigma\log\rho_\text{lin}(t,\vx)\big)\big)} \notag
\end{align}

\noindent 
It was unnecessary to modify the $U_a$ and $U_r$ PDEs since they are linear. Source function \eqref{E:sourcecomp} was computed symbolically from linear solutions (\ref{E:CTsolnrho1D}-\ref{E:CTsolnUr1D}) or (\ref{E:CTsolnrho2D}-\ref{E:CTsolnUr2D})  and converted to \texttt{sympy} expressions with Mathematica \cite{WolframResearch2019}. \texttt{sympy} is a computer algebra package for the programming language \texttt{python}. 

\begin{table}
    \begingroup\centering
    \begin{tabular}{S|S|S|S|S}
        \multicolumn{5}{c}{\textbf{A. Time step size series, one dimensional}} \\
         & 
         \multicolumn{2}{l|}{\text{$\norm{error}$ (\si{cm^{-1}})}} & \multicolumn{2}{|l}{\text{convergence rate}}  \\
         \text{$\Delta t$ (\si{s})}& $L^{2}$ & $L^\infty$ & $L^{2}$ & $L^\infty$ \\
         \hline
         4 & 0.0100 & 0.0220 & -0.064 & -0.054\\
         8 & 0.0096 & 0.0212 & 0.210 & 0.237 \\
         16 & 0.0111 & 0.0250 & 1.16 & 1.14 \\
         32 & 0.0247 & 0.0552 & 2.28 & 2.26 \\
         64 & 0.120 & 0.265 & 2.95 & 2.93 \\
         128 & 0.930 & 2.02 & 3.08 & 3.07 \\
         256 & 7.88 & 17.0 & & \\
         \hline
    \end{tabular}
    \endgroup
    \vspace{1em}\newline
    \begingroup\centering
    \begin{tabular}{S|S|S|S|S}
        \multicolumn{5}{c}{\textbf{B. Spatial point distance series, one dimensional}} \\
         & 
         \multicolumn{2}{l|}{\text{$\norm{error}$ (\si{cm^{-1}})}} & \multicolumn{2}{|l}{\text{convergence rate}}  \\
         \text{$\Delta x$ (\si{cm})}& $L^{2}$ & $L^\infty$ & $L^{2}$ & $L^\infty$ \\
         \hline
         {1/1024} & 0.0086 & 0.0196 & 0.216 & 0.162 \\
         {1/512} & 0.0100 & 0.0220 & 1.28 & 1.22 \\
         {1/256} & 0.0244 & 0.0514 & 3.43 & 3.33\\
         {1/128} & 0.263 & 0.516 & 3.95 & 3.95 \\
         {1/64} & 4.07 & 7.97 & & \\
         \hline
    \end{tabular}
    \endgroup
    \caption{Convergence test results, one spatial dimensional}
    \label{T:convergence1D}
    Eqs. (\ref{E:convrhoPDE}, \ref{E:UaPDE}, \ref{E:UrPDE}) were solved numerically from $t = $ \SIrange{0}{8192}{s} on $x\in\Omega=[0,1]$. In this time the amplitude of the sinusoid $\delta\rho$ \eqref{E:CTsolndrho1D} grew from $a_\rho = 1$ to $a_\rho e^{8192\lambda} \approx 2505$. \textbf{A} shows the results of varying the time step size from \numrange{4}{256} \si{s} (with a fixed spatial point distance of $\Delta x = $\SI{1/512}{cm}). \textbf{B} shows the results of varying the spatial point distance from $1/1024$ to $1/64$ \si{cm} (with a fixed time step of \SI{4}{s}). The error in $\rho$ at the final time point was calculated as the difference between the numerical result and exact result \eqref{E:CTsolnrho1D}. $L^2$ and $L^\infty$ norms of the error are tabulated. Convergence rate is calculated between consecutive rows as $\log(\norm{\text{error}_1}/\norm{\text{error}_2})/\log(h_1/h_2)$, with $h$ being either $\Delta t$ or $\Delta x$, as appropriate. The mean of $\rho$ was \SI{9000}{cm^{-1}} in all cases. Thus the relative error is about $1/9000$ times the error shown---e.g. $0.0100/9000\approx$ \num{1.1e-6} for $\Delta t = $ \SI{4}{s}, $\Delta x = 1/512$ \si{cm} in one dimension. Errors in $U_a$ and $U_r$ (not shown) were smaller but otherwise behaved similarly. All numerical solutions used the \petsc Rosenbrock-W solver \texttt{ra34pw2} (nominally an order 3 method), and fourth-order approximations for the spatial derivatives.
\end{table}

\begin{table}
    \begingroup\centering
    \begin{tabular}{S|S|S|S|S}
        \multicolumn{5}{c}{\textbf{A. Time step size series, two dimensional}} \\
         & 
         \multicolumn{2}{l|}{\text{$\norm{error}$ (\si{cm^{-1}})}} & \multicolumn{2}{|l}{\text{convergence rate}}  \\
         \text{$\Delta t$ (\si{s})}& $L^{2}$ & $L^\infty$ & $L^{2}$ & $L^\infty$ \\
         \hline
         4 & 0.0072 & 0.0469 & 2.01 & 2.20\\
         8 & 0.0290 & 0.215 & 0.692 & 0.627 \\
         16 & 0.0468 & 0.332 & -1.57 & -2.21 \\
         32 & 0.0158 & 0.0716 & 3.40 & 3.48 \\
         64 & 0.167 & 0.800 & 1.53 & 1.28 \\
         128 & 0.483 & 1.94 & 3.14 & 3.21 \\
         256 & 4.27 & 18.0 & & \\
         \hline
    \end{tabular}
    \endgroup
    \vspace{1em}\newline
    \begingroup\centering
    \newline
    \begin{tabular}{S|S|S|S|S}
        \multicolumn{5}{c}{\textbf{B. Spatial point distance series, two dimensional}} \\
         & 
         \multicolumn{2}{l|}{\text{$\norm{error}$ (\si{cm^{-1}})}} & \multicolumn{2}{|l}{\text{convergence rate}}  \\
         \text{$\Delta x$ (\si{cm})}& $L^{2}$ & $L^\infty$ & $L^{2}$ & $L^\infty$ \\
         \hline
         {$\sqrt{3}/2048$} & 0.0246 & 0.178 & -1.77 & -1.92 \\
         {$\sqrt{3}/1024$} & 0.0072 & 0.0469 & 0.103 & -0.172 \\
         {$\sqrt{3}/512$} & 0.0077 & 0.0416 & 2.84 & 2.26\\
         {$\sqrt{3}/256$} & 0.0555 & 0.199 & 3.92 & 3.86 \\
         {$\sqrt{3}/128$} & 0.838 & 2.88 & & \\
         \hline
    \end{tabular}
    \endgroup
    \caption{Convergence test results, two spatial dimensions}
    \label{T:convergence2D}
    Eqs. (\ref{E:convrhoPDE}, \ref{E:UaPDE}, \ref{E:UrPDE}) were solved numerically from $t = $ \SIrange{0}{8192}{s} on $(x,y)\in\Omega=[0,\sqrt{3}/2]\times [0,1/2]$. In this time the amplitude of the sinusoid $\delta\rho$ \eqref{E:CTsolndrho2D} grew from $a_\rho = 1$ to $a_\rho e^{8192\lambda} \approx 2505$. \textbf{A} shows the results of varying the time step size from \numrange{4}{256} \si{s} (with a fixed spatial point distance of $\Delta x = \sqrt{3}/1024$ \si{cm}). \textbf{B} shows the results of varying the spatial point distance from $\sqrt{3}/2048$ to $\sqrt{3}/128$ \si{cm} (with a fixed time step of \SI{4}{s}). In all cases $\Delta y = \Delta x\times 64\sqrt{3}/111$. Errors and convergence rates calculated as in Table \ref{T:convergence1D}.
\end{table}

\subsection*{Software}

Software developed for this work is available from \texttt{https://github.com/leonavery/KSFD}. Morpheus \cite{Starruss2014} implementations of the cellular Potts models are available from \texttt{https://github.com/leonavery/worm-CPM}. The \texttt{README} file for \texttt{worm-CPM} discusses the suitability of the cellular Potts model for worms in some detail.

\subsection*{Supplemental figures}
%Re-run figures:

%attractant only
%options143
%AttractantOnly_reseed.ai,.png

\begin{figure}[!ht]
    \hrule
    \vspace{12pt}
    \begin{center}
    \includegraphics[width=5in]{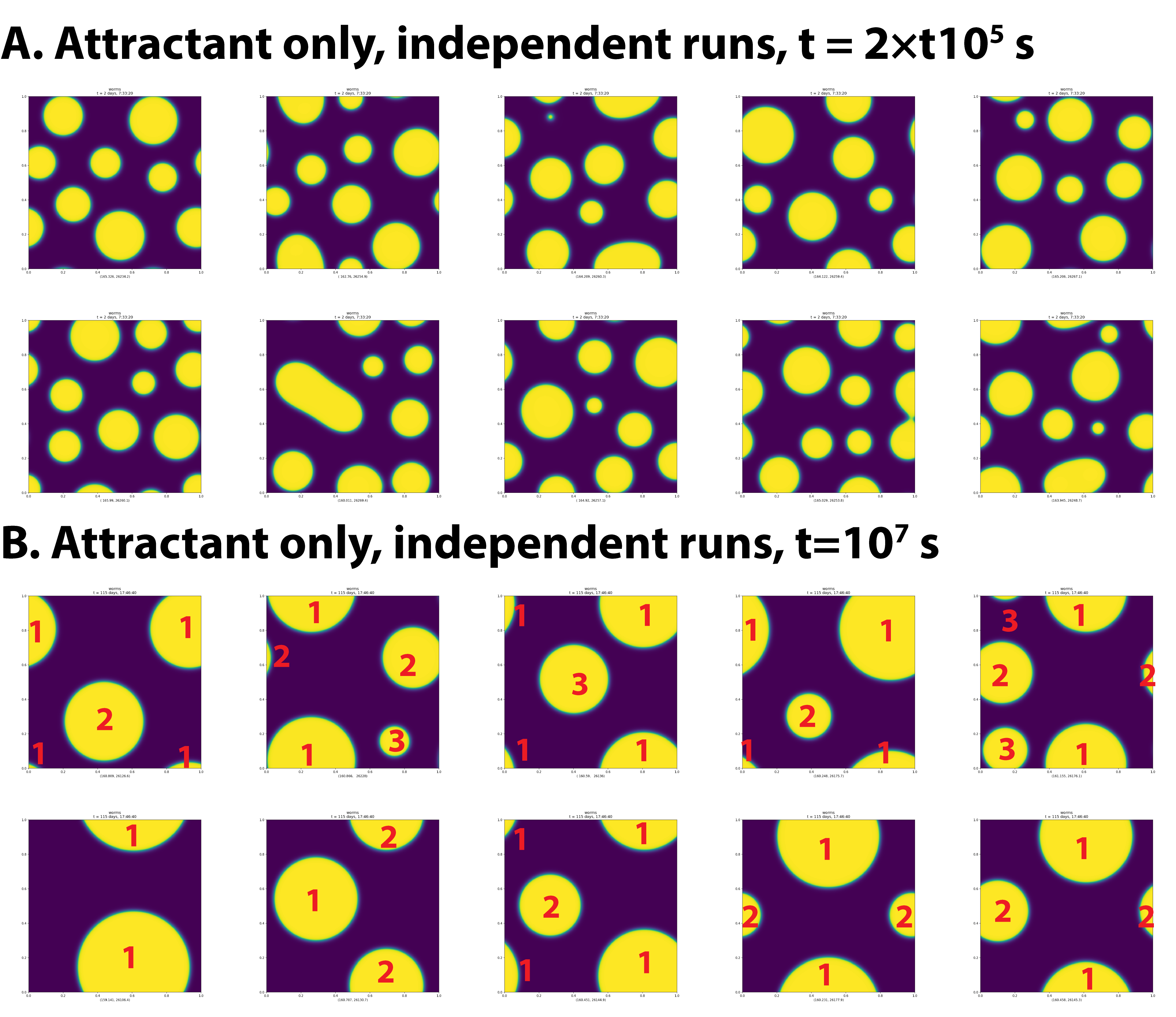}
    \end{center}

    \vspace{6pt}

    \caption{Attractant-only simulation reruns}
    \textbf{A.} These ten images reproduce the numerical experiment of Figure \ref{F:fig3.periodic}B---simulation of the attractant-only model in two dimensions---but with different pseudorandom noise in the initial condition. Only worm density $\rho(x,y)$ at $t = $ \SI{200000}{s} (2 days, 7:33:20) is shown. \textbf{B.} Like \textbf{A}, but at $t = $ \SI{1e7}{s} (115 days, 17:46:40). These images correspond one-to-one to the images in \textbf{A}. The number of aggregates in these panels ranges from one to three, although a single aggregate may appear in as many as four pieces because of the periodic boundary conditions. To ease the identification of aggregates, the aggregate to which each piece belongs is identified by a red number.
    \label{F:fig3.AttractantReruns}
    \vspace{6pt}
    \hrule
\end{figure}

\begin{figure}[!ht]
    \hrule
    \vspace{12pt}
    \begin{center}
    \includegraphics[width=5in]{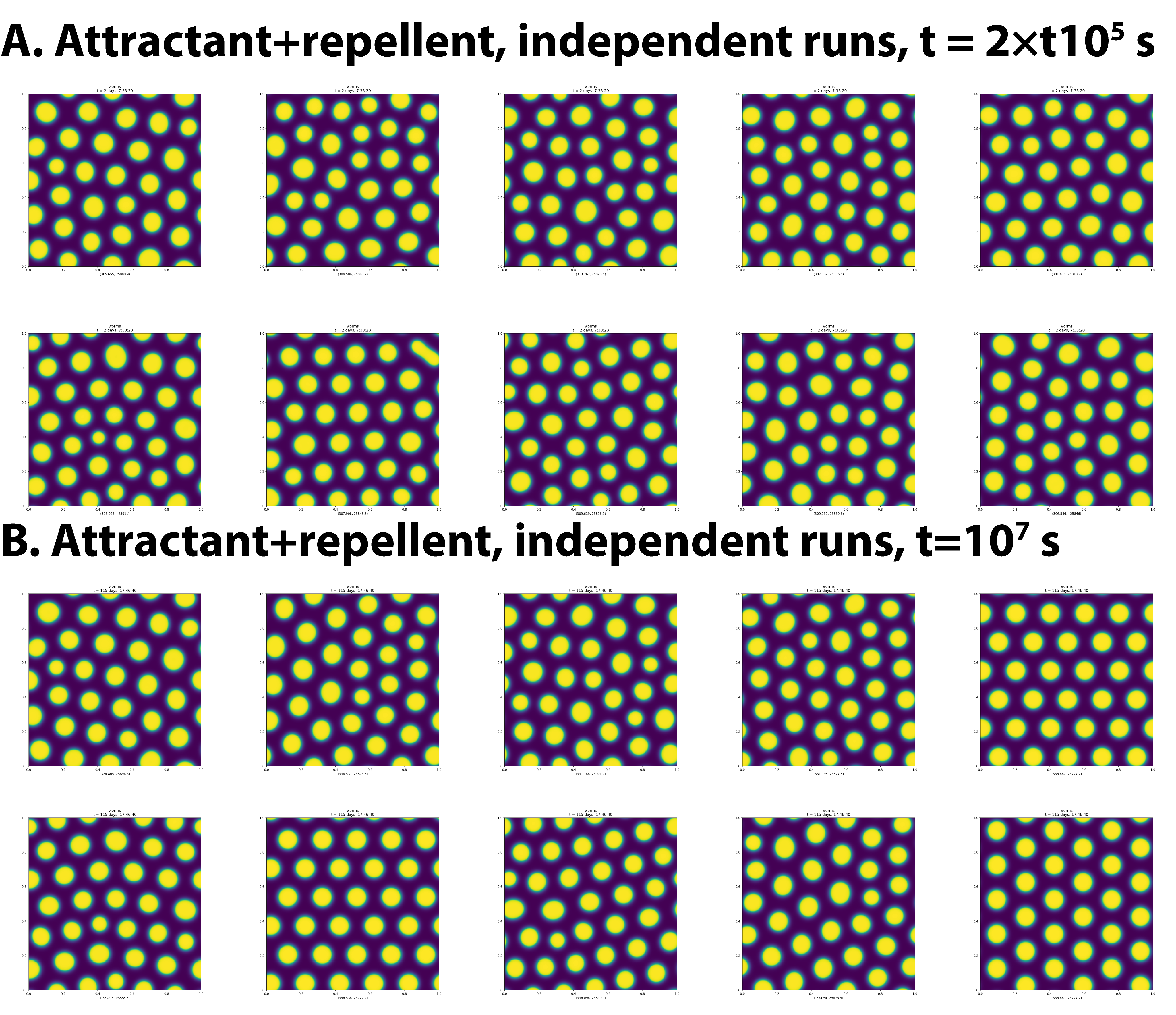}
    \end{center}

    \vspace{6pt}

    \caption{Attractant+repellent simulation reruns}
    \textbf{A.} These ten images reproduce the numerical experiment of Fig \ref{F:fig3.repellent}B---simulation of the attractant+repellent model in two dimensions---but with different pseudorandom noise in the initial condition. Only worm density $\rho(x,y)$ at $t = $ \SI{200000}{s} (2 days, 7:33:20) is shown. \textbf{B.} Like \textbf{A}, but at $t = $ \SI{1e7}{s} (115 days, 17:46:40). These images correspond one-to-one to the images in \textbf{A}.
    \label{F:fig3.AttractantRepellentReruns}
    \vspace{6pt}
    \hrule
\end{figure}

\begin{figure}
    \hrule
    \vspace{12pt}
    \begin{center}
    \includegraphics[width=5in]{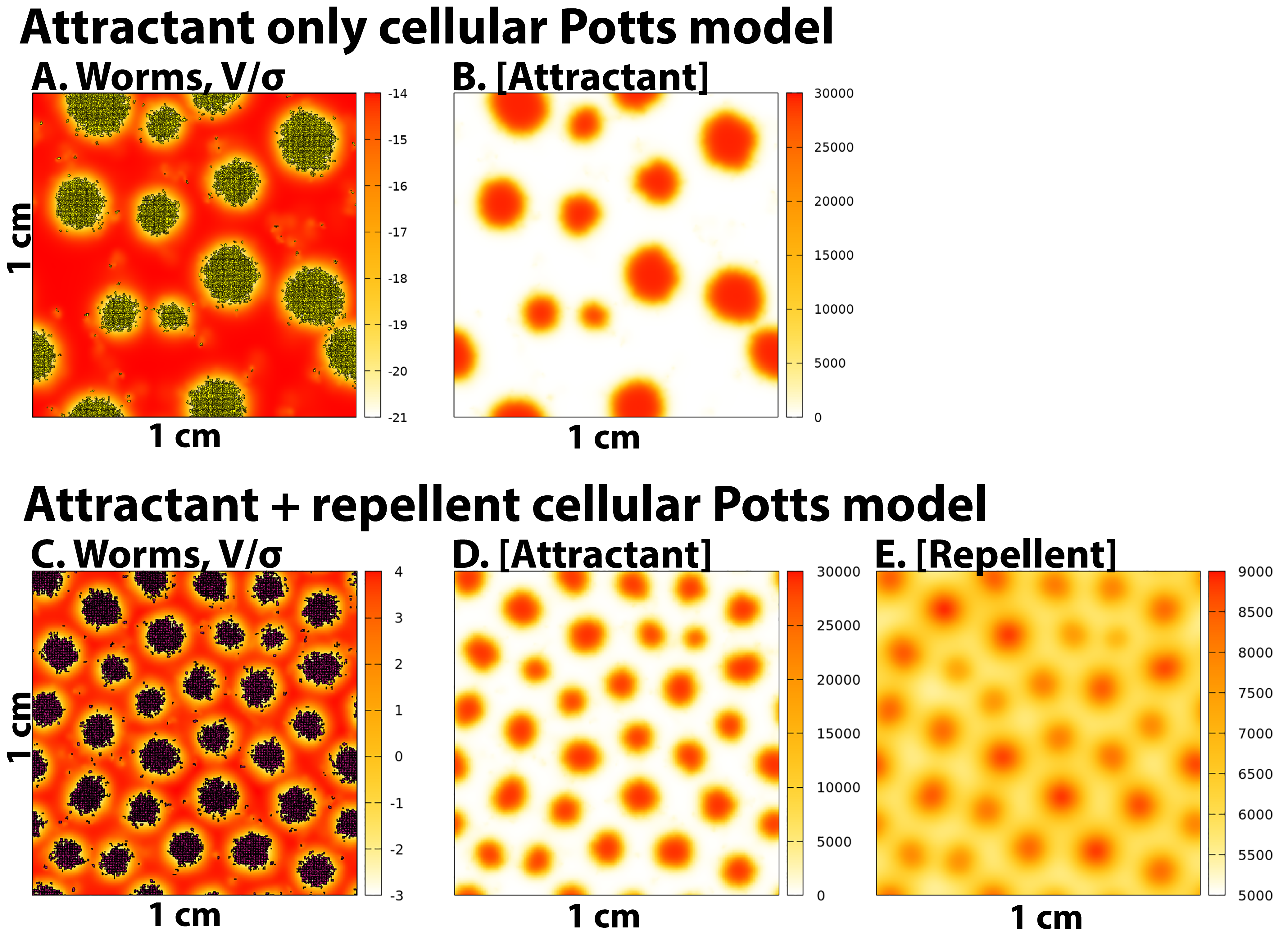}
    \end{center}

    \vspace{6pt}

    \caption{Cellular Potts model simulations}
    \textbf{A, B} These two images show the results of an individual-based cellular Potts model simulation of the Attractant-only model in two dimensions, and are meant to be compared to Fig \ref{F:fig3.periodic}C,D. \textbf{D-F} show the results of a cellular Potts model simulation of the attractant+repellent model in two dimensions and can be compared to Fig \ref{F:fig3.repellent}C,D,E. Because there is no simple relationship between the parameters of the PDE model and the cellular Potts model, we do not expect precise quantitative agreement, even on a statistical basis.
    \label{F:fig3.CPM}
    \vspace{6pt}
    \hrule
\end{figure}
%attractant-only: worm5g.xml
%attractant+repellent: worm6c.xml
%CPM.ai,.png

\subsection*{Supplemental data}

The following data file is provided:

\begin{description}
    \item[C.elegans microarray results\_020615\_CID.xlsx] Microarray expression profiling results. 
\end{description}
\subsection*{Videos}
The following video files are provided. 

Avery, Leon (2021), “Avery\_L1agg2”, Mendeley Data, V3, doi: 10.17632/r5v772ftcs.3

\begin{description}
    \item[N2\_5e5\_washed.avi] This video shows the time course of aggregation after \num{500000} starved L1s were pipetted onto the center of a petri plate. The recording covers \SI{720}{\minute}. There is one frame per minute of real time, and the playback rate is \SI{7}{s^{-1}}. 
    \item[options138a.mp4] Numerical solution of attractant-only model in one dimension. This video corresponds to Fig \ref{F:fig3.periodic}A. This and all following videos are 200s long at \SI{15}{s^{-1}}. Time is displayed as ``days, H:MM::SS''. Time ranges from \SI{0}{s} to \SI{200000}{s} (2 days, 7:33:20). The two numbers below each panel are the minimum and maximum of the plotted field.
    \item[options139.mp4] Numerical solution of attractant-only model in two dimensions. Corresponds to Fig \ref{F:fig3.periodic}B.
    \item[options140a.mp4] Numerical solution of attractant+repellent model in one dimension. Corresponds to Fig \ref{F:fig3.repellent}A.
    \item[options141.mp4] Numerical solution of attractant+repellent model in two dimensions. Corresponds to Fig \ref{F:fig3.repellent}B.
    \item[options157.mp4] Numerical solution of attractant+repellent model in two dimensions on a \SI{6x6}{cm} domain, with most worms initially placed in the center. Corresponds to Fig \ref{F:fullscale}.
    \item[worm5g.mp4] Numerical solution of attractant-only cellular Potts model in two dimensions. Corresponds to Fig \ref{F:fig3.CPM}A,B.
    \item[worm6c.mp4] Numerical solution of attractant+repellent cellular Potts model in two dimensions. Corresponds to Fig \ref{F:fig3.CPM}C,D,E.
\end{description}

%videos:
%1d attractant only: options138a.mp4
%2d attractant only: options139.mp4
%1d attractant + repellent: options140a.mp4
%2d attractant + repellent: options141.mp4
%full-scale: options157.mp4

\end{document}